\documentclass[12pt]{article}
\usepackage{latexsym,amsmath,amsfonts,amssymb}
\usepackage[latin1]{inputenc}
\usepackage[pdftex]{graphicx}

\usepackage{euscript,mathrsfs}
\usepackage{fullpage}
\usepackage{color}
\usepackage{bbm}
\usepackage[american]{babel}

\makeatletter \@addtoreset{equation}{section}

\makeatletter\renewcommand\section{\@startsection {section}{1}{\z@}%
                                   {-3.5ex \@plus -1ex \@minus -.2ex}
                                   {2.3ex \@plus.2ex}%
                                   {\normalfont\large\bfseries}}
\renewcommand\subsection{\@startsection{subsection}{2}{\z@}%
                                     {-3.25ex\@plus -1ex \@minus -.2ex}%
                                     {1.5ex \@plus .2ex}%
                                     {\normalfont\bfseries}}

\renewcommand{\baselinestretch}{1.2}
\newcommand{\be}{\begin{equation}}
\newcommand{\ee}{\end{equation}}
\newcommand{\bea}{\begin{eqnarray}} 
\newcommand{\eea}{\end{eqnarray}}

\newcommand{\comment}[1]{}

\newcommand{\p}{\partial}

\newcommand\e{\epsilon}

\newcommand{\C}{\mathbb{C}}
\newcommand{\R}{\mathbb{R}}
\newcommand{\Z}{\mathbb{Z}}

\newcommand{\RP}{\mathbb{RP}}

\renewcommand{\H}{{\cal H}}

\newcommand{\M}{{\cal M}}
\newcommand{\scri}{{\cal I}}

\begin{document}

\begin{titlepage}
\vspace{6cm}
\vfil\

\begin{center}
{\LARGE A de Sitter Farey Tail} 

\vspace{6mm}
{Alejandra Castro\footnote{e-mail: {\tt acastro@physics.mcgill.ca}}, 
 Nima Lashkari\footnote{e-mail: {\tt lashkari@physics.mcgill.ca}}, 
 \&
Alexander Maloney\footnote{e-mail: {\tt maloney@physics.mcgill.ca}}
}\vspace{6.0mm}\\
\bigskip
{\it 
McGill Physics Department, 3600 rue University, Montreal, QC H3A 2T8, Canada}
\medskip
\vfil

\end{center}
\setcounter{footnote}{0}

\begin{abstract}
\noindent
We consider quantum Einstein gravity in three dimensional de Sitter space.  The Euclidean path integral is formulated as a sum over geometries, including both perturbative loop corrections and non-perturbative instanton corrections coming from geometries with non-trivial topology.    These non-trivial geometries have a natural physical interpretation.  Conventional wisdom states that the sphere is the unique Euclidean continuation of de Sitter space.  However, when considering physics only in the causal patch of a single observer other Euclidean geometries, in this case lens spaces, contribute to physical observables.  This induces quantum gravitational effects which lead to deviations from the standard thermal behaviour obtained by analytic continuation from the three sphere.  The sum over these geometries can be formulated as a sum over cosets of the modular group; this is the de Sitter analog of the celebrated ``black hole Farey tail."  We compute the vacuum partition function including the sum over these geometries.  Perturbative quantum corrections are computed to all orders in perturbation theory using the relationship between Einstein gravity and Chern-Simons theory, which is checked explicitly at tree and one-loop level using heat kernel techniques.  The vacuum partition function, including all instanton and perturbative corrections, is shown to diverge in a way which can not be regulated using standard field theory techniques.
\end{abstract}
\vspace{0.5in}

\end{titlepage}
\newpage

\renewcommand{\baselinestretch}{1.1}  
\renewcommand{\arraystretch}{1.5}

\tableofcontents

\section{Introduction}

As the maximally symmetric solution of general relativity with a positive cosmological constant,  de Sitter space is the natural starting point for the study of quantum cosmology.  Despite notable efforts  it is not known precisely how to define a theory of quantum gravity in eternal de Sitter space or even whether such a theory exists (see e.g. \cite{Witten:2001kn, Strominger:2001pn,Maldacena:2002vr,Goheer:2002vf,Kachru:2003aw}).  Such a theory would presumably answer several important questions, including:
\begin{itemize}
\item
What are the appropriate observables for eternal inflation?  
\item 
What is the origin and interpretation of the Bekenstein-Hawking entropy of a cosmological horizon?  
\item
How does quantum gravity alter the physics of observers in a de Sitter universe?
\end{itemize}
We will not attempt to provide a full answer to these questions here; rather we will describe a series of explicit computations which will shed some light on the third question. We will return to the second at the end of this paper. 

Our focus here is on three dimensional de Sitter gravity, where the computations are simple and quantum corrections can be computed systematically. We will make precise the notion of a path integral as a sum over all smooth geometries in Euclidean signature, and discuss the physical interpretation of this path integral.   Although we will not use explicitly any notions from string theory or holography, our approach is inspired by the corresponding analysis in AdS$_3$ based on the AdS/CFT correspondence.

\subsection{de Sitter Space and Thermality}

Conventional wisdom states that de Sitter space is thermal, in the sense that the de Sitter horizon emits a bath of Hawking radiation at a fixed temperature.  We will argue that this statement must be modified once quantum gravity effects are taken into account.  In  three dimensional de Sitter gravity the deviations from the standard canonical ensemble can be computed exactly, but we expect a similar statement to be true in higher dimensions as well.

We begin by first recalling why de Sitter space is thermal.  
For field theory in a fixed curved background, unlike in flat Minkowski space, there is no unique choice of vacuum.  The typical choice of vacuum -- often referred to as the Hartle-Hawking, or Euclidean, vacuum state  \cite{Hartle:1983ai} -- is defined by Wick rotation from Lorentzian to Euclidean signature.  More precisely, the ground state wave functional, viewed as a function of field configurations on a constant time slice, is computed by performing a Euclidean path integral with specified data on the constant time slice.  The natural Euclidean continuation of Lorentzian de Sitter space is the sphere $S^3$; in this continuation the time coordinate of a static observer becomes an angular coordinate of the sphere.  Correlation functions computed in this state are found by analytic continuation of correlation functions on $S^3$.  These correlators are periodic in Euclidean time and obey the KMS conditions.  So correlation functions of field operators are evaluated in a canonical ensemble state at fixed temperature.

In this paper we will assume that the definition of the Hartle-Hawking state in terms of Euclidean functional integral is, to the extent that it can be made precise, correct.  This means that once quantum gravitational effects are included the vacuum state will include contributions from other geometries.  In particular, other solutions to the Euclidean equations of motion will appear as saddle point contributions.  In three dimensions these other saddle points are easy to describe; they are quotients $S^3/\Gamma$ of the three sphere by a discrete subgroup $\Gamma$ of $SO(4)$.  In this paper we focus on a particularly simple class of such geometries of the form $S^3/\Z_p$. 
These spaces, known as lens spaces, have a very simple physical interpretation.  Correlation functions which are defined by analytic continuation from a lens space do not describe a canonical ensemble state at fixed temperature.  Rather, they describe a grand canonical ensemble state at fixed temperature and angular potential.  In such a state the de Sitter horizon emits Hawking radiation at a fixed angular potential, much like a rotating black hole.

The full Hartle-Hawking state includes a sum over the quotients of $S^3$. At the classical level, each geometry is weighted by its Euclidean action. The leading contribution will be the familiar thermal state coming from the dominant $S^3$ saddle.  The quotients give subleading contributions which are suppressed by terms that are exponentially large in the de Sitter entropy; they vanish in the classical limit.  Nevertheless their effects can be computed and the full partition function includes a sum over lens spaces.  In fact, this sum over lens spaces has an elegant mathematical interpretation as a sum over the modular group $SL(2,\Z)$.   This is very reminiscent of the ``black hole Farey tail" of \cite{Dijkgraaf:2000fq}.  In that case the partition function of AdS${}_3$ gravity at finite temperature is interpreted as a sum over the modular group.  The sum over $SL(2,\Z)$ was a sum  over all three dimensional geometries which ``fill in'' a $T^2$ at the conformal boundary of space-time  \cite{Maldacena:1998bw,Dijkgraaf:2000fq,Kraus:2006nb,Maloney:2007ud}.  In the de Sitter case, this sum has a similar interpretation as a sum over ways of filling in
a $T^2$ at the Euclidean horizon.  Thus our ``de Sitter Farey tail'' provides a construction of Hartle-Hawking state as a sum over geometries related by modular transformations. 

\subsection{The Partition Function of de Sitter Gravity}

The geometries described above all contribute to the partition function of de Sitter gravity in Euclidean signature.  Formally, 
this partition function should be regarded as a path integral over Euclidean metrics
\be\label{Zdef}
Z = \int {\cal D}g~ e^{-S[g] }~.
\ee
The Hartle-Hawking state is an integral over metrics with fixed boundary conditions on a space-like slice.
The path integral in equation (\ref{Zdef}) is a sum over all compact metrics and is interpreted as the norm of the Hartle-Hawking state.
In the present work we will consider the case of ``pure" Einstein gravity with only metric degrees of freedom.

Of course, there are very few cases where we know how to make precise sense of a sum over geometries of the form (\ref{Zdef}).  When the cosmological constant is negative, the integral is defined over metrics with fixed conformal structure at the boundary and the result can be identified with the partition function of a conformal field theory.  This allows one to make a certain amount of progress.  It turns out that when the cosmological constant is positive  there is an alternate technique -- not based on AdS/CFT -- which provides a precise, calculable definition of the partition function (\ref{Zdef}).

To begin, we first write down the saddle point approximation to the partition sum (\ref{Zdef})
\be\label{Zsaddle}
Z = \sum_{g_c} e^{-k S^0[g_c] + S^1[g_c]+ {1\over k} S^2[g_c]+\dots}~.
\ee
Here the sum is over all classical solutions $g_c$ to the Euclidean equations of motion, and $S^{i}[g_c]$ denotes the quantum correction to the action at $i^{th}$ order in perturbation theory.  We have extracted explicitly the dimensionless coupling constant $k$, which in the present case is equal to the de Sitter radius in Planck units.  In general the expression (\ref{Zsaddle}) will only be an approximation to the full path integral.  However, this approximation is expected to become exact if we can accomplish the following two tasks:
\begin{itemize}
\item
Identify the infinite set of classical solutions $g_c$
\item
Compute the infinite series of subleading corrections $S^i$ around each classical saddle
\end{itemize}
We will perform both of these computations.

The identification of the classical saddles $g_c$ is easy.  They are the quotients $S^3/\Gamma$, which can be ennummerated  and described explicitly.  In principle, a given saddle may or may not contribute to the path integral --  to answer this question one must define the Lorentzian path integral precisely, rotate the integral to Euclidean signature and determine which saddles lie on the contour of stationary phase.  We will not attempt to do so in this paper; instead we focus on the lens spaces $L(p,q)$, whose inclusion in the path integral can be motivated on physical grounds.
We will leave the question of more complicated quotients to future work.\footnote{We will, however,  compute the classical sum over geometries for all possible quotients   $S^3/\Gamma$ in an appendix.}

In order to compute the series of perturbative corrections, we will use the relationship between three dimensional Einstein gravity and Chern-Simons theory \cite{Witten:1988hc}.   We emphasize that we will {\it not} attempt to identify the gravitational path integral with that of a Chern-Simons theory.  Indeed, the sum over saddle points takes a very different form in these two theories.  For example, our gravitational path integral (\ref{Zdef})  involves a sum over geometries with different topology, whereas the Chern-Simons partition function is a sum over flat connections on a space of fixed topology.  However, at the level of perturbation theory, the rewriting of gravitational degrees of freedom in terms of a gauge connection is straightforward.  Thus one can use Chern-Simons theory as an efficient computational tool to extract perturbative corrections to a given classical saddle point.  In the Chern-Simons formulation there is a systematic perturbative expansion \cite{Witten:1988hc, Witten:1988hf, Achucarro:1987vz, Achucarro:1989gm,Gukov:2003na}, and remarkably the partition function on a  lens space is known to all orders  \cite{Jeffrey:1992tk}.   This will allow us to identify the infinite set of subleading perturbative corrections in (\ref{Zsaddle}).

We note that the use of the Chern-Simons formulation as a computational tool in perturbative gravity is somewhat delicate.  Previous related efforts include \cite{Carlip:1992wg,Guadagnini:1995wv,Banados:1998tb, Park:1998yw, Govindarajan:2002ry}.  For example, in the case of a negative cosmological constant both the space-time and the Chern-Simons gauge group are non-compact, which makes the gauge theory formulation subtle.  Thankfully, these subtleties do not occur in the present case.  Euclidean gravity with a positive cosmological constant is related to $SO(4)$ Chern-Simons theory on a compact space $S^3/\Gamma$, so the Chern-Simons formulation is straightforward.

We will devote some time to a precise matching between the Chern-Simons and gravitational results.  At the classical level, the comparison is simple; the Einstein action is related to a particular Chern-Simons invariant  \cite{Witten:1988hc}.  We will also check that Einstein gravity is equal to Chern-Simons theory at the one-loop level as well.  This is considerably more involved.  We perform  a direct computation of the one-loop determinant for gravity on a lens space using heat kernel techniques, and show that this is exactly equal to the corresponding Chern-Simons result.  To our knowledge this is the first time such a check has been done.  Given this check at the tree and one-loop level, we can then confidently apply the Chern-Simons result at all orders in perturbation theory.

The result is that the partition function (\ref{Zdef}) is divergent.
This is not a surprise, since it involves a sum over quotients $S^3/\Gamma$.  As the order of the group $\Gamma$ goes to infinity, we find an infinite number of saddles whose classical actions approach zero.  Indeed, this same divergence appeared in the case of a negative cosmological constant \cite{Dijkgraaf:2000fq}.  In that case it was necessary to apply a regularization scheme which rendered the sum finite and provided a match with the CFT.  In fact, the sum over geometries in that case could be regulated only for certain special values of the coupling constants, for example when the central charges satisfy ${1\over 24} (c_L-c_R) \in \Z$.   The obvious questions is whether the same is true in the case of gravity with a positive cosmological constant.  In particular, we ask whether an appropriate regularization will render the quantum path integral for de Sitter gravity finite.  Optimistically, this would impose a constraint on the coupling constant of the theory -- i.e. the cosmological constant -- required for quantum mechanical consistency.  We will conclude that this is not the case for pure Einstein gravity in de Sitter space.
The perturbative corrections to the effective action for de Sitter gravity make the sum over geometries converge more rapidly, but there is still a divergent piece even after regularization.  Unlike the case of a negative cosmological constant, there appears to be no way of regulating the divergence appearing in the sum over geometries.  We will discuss possible implications and interpretations of this result.

\subsection{Overview}
In section 2 we describe the geometry of the lens spaces and demonstrate that they construct states for static patch observers in a grand canonical ensemble.  In section 3 we  describe the Euclidean path integral in the saddle point approximation and perform the classical sum over lens spaces.  We will also discuss the regularization of this sum.   In section 4 we compute the one-loop correction to this classical sum, using both heat kernel and Chern-Simons techniques.     In section 5 we compute the sum using the result at all orders in perturbation theory. We end in section 6 with a discussion of open issues and speculations regarding the entropy of de Sitter space.   

In appendix A we describe the regularized sum over all Euclidean saddles $S^3/\Gamma$.  We summarize some formulae relevant for the computations of one-loop determinants and zeta function regularization in appendices B \& C.

\section{Lens Spaces and the Static Patch}\label{sec:2}

In this section we motivate the inclusion of lens spaces in the path integral of de Sitter gravity and describe their physical interpretation.

\subsection{The Static Patch and the Hartle-Hawking State}

We start by considering the physics of a timelike observer in dS${}_3$.  Such an observer is in causal contact with only a portion of the full de Sitter geometry.  This region is known as the static patch (or causal diamond) associated with the observer.  The metric on the static patch of de Sitter space can be written as
\be\label{lmetric}
{ds^2\over \ell^2} = dr^2 - \cos^2 r dt^2 + \sin^2 r d\phi^2~. 
\ee
Here $\ell$ is the curvature radius and $\phi$ is periodically identified 
\be
\phi \sim \phi + 2 \pi n~~~~~~\forall n \in \Z~.
\ee
In these coordinates the observer is located at $r=0$.  The boundary of the static patch is the cosmological horizon, which in these coordinates is located at $r=\pi/2$ (see figure 1).  We will use units $\ell=1$.

\begin{figure}\label{dsfig}
\centering
{\includegraphics[width=0.4\textwidth]{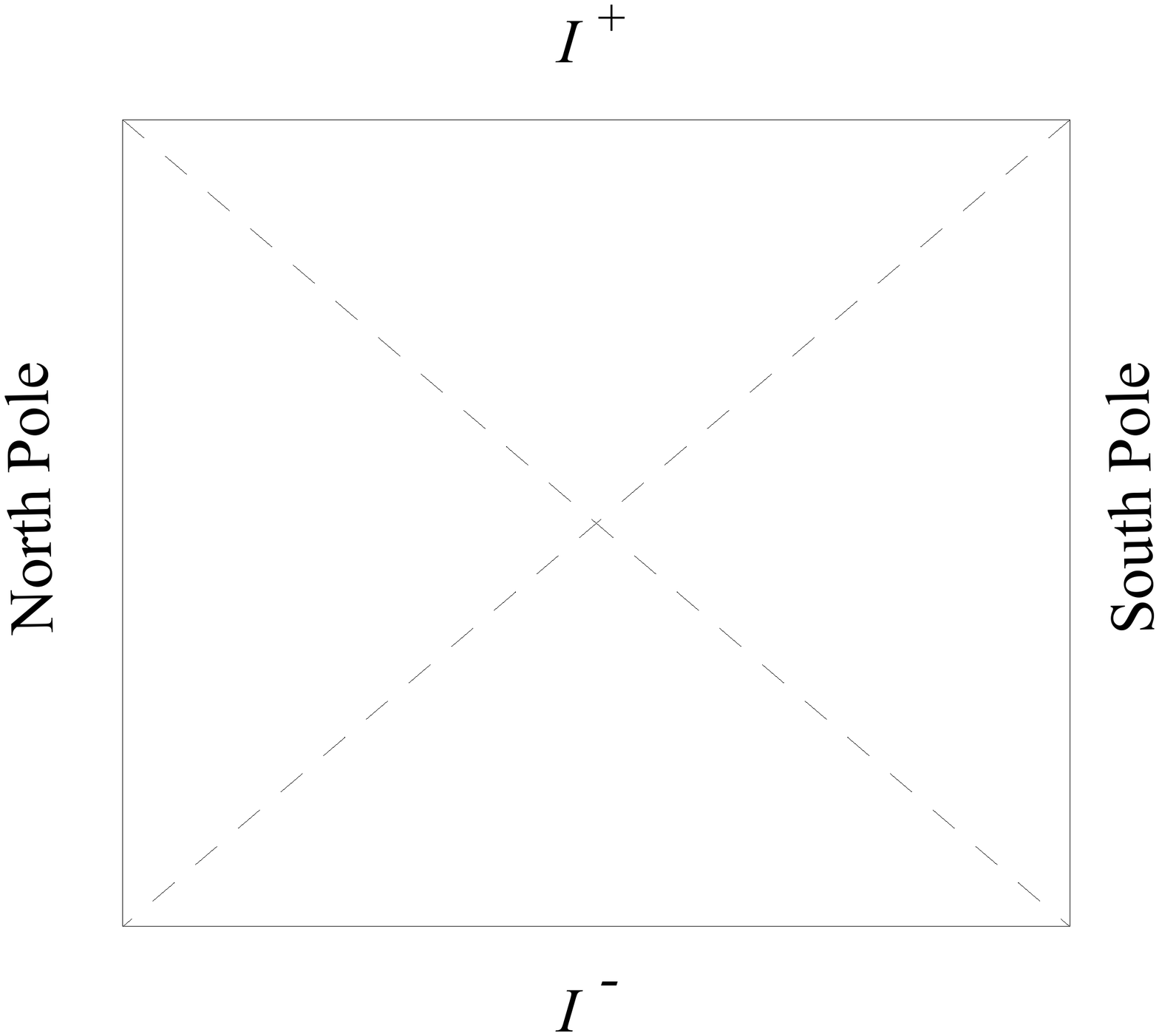}
~~~~~~~~~~~~
  \includegraphics[width=0.4\textwidth]{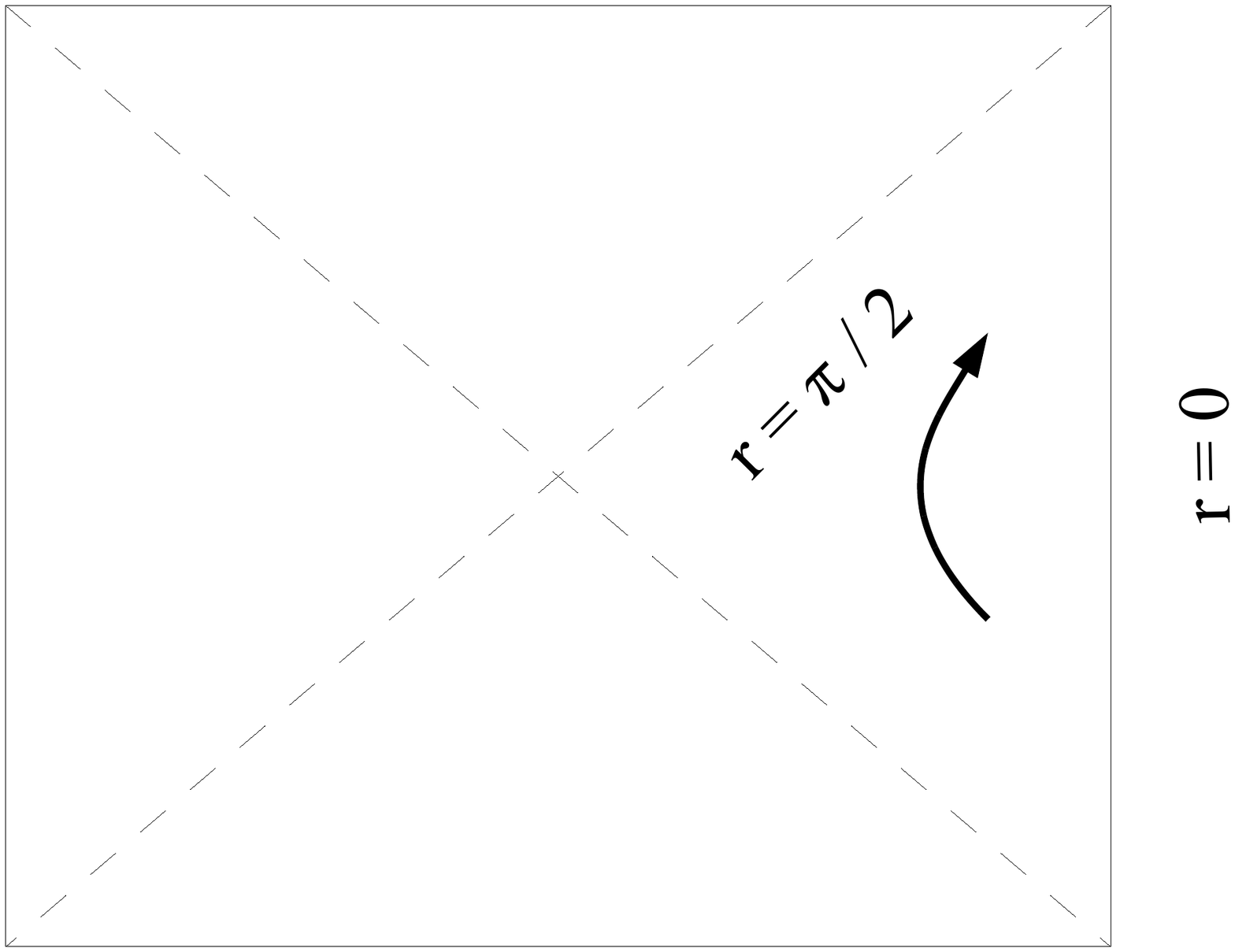}}
  \caption{The causal diagram of de Sitter space, suppressing the angular $\phi$ coordinate.  Horizontal slices of this Penrose diagram are spheres $S^2$, with north and south poles given by vertical lines on the left and right, respectively.  The static patch associated with an observer at the south pole is the wedge shaped region on the right.  In our coordinates the observer is at $r=0$ and the cosmological horizon is at $r=\pi/2$.  On this horizon the timelike Killing vector (denoted by an arrow) becomes null.  }
\end{figure}

In this coordinate system the metric is static and axially symmetric.  These symmetries are generated by the Killing vectors
\be\label{hj}
H = i \p_t~,~~~~~~J = i \p_\phi~.
\ee 
These vectors define a notion of energy and angular momentum associated with this patch.  We note that the timelike Killing vector $H$ becomes null at the horizon at $r=\pi/2$.  Indeed, de Sitter space does not possess a globally timelike Killing vector.  Correspondingly, there is no global notion of conserved energy in de Sitter space.  The best one can do is consider charges of the sort defined in (\ref{hj}) associated to a particular observer.

Before turning to the physics of quantum gravity, it is useful to first consider free quantum field theory in a fixed de Sitter background.  Restricting our attention to the static patch, it is straightforward to construct phase space charges $H$ and $J$ which generate the Killing symmetries (\ref{hj}).  Upon quantization, these will become operators acting on the field theory Hilbert space.  In free field theory this can be done completely explicitly and the Hilbert space can be organized into states of fixed energy and angular momentum. Since $\phi$  is periodically identified the charge $J=i \p_\phi$ will take integer values.

In order to define quantum field theory in the de Sitter background it is necessary to choose a vacuum state.  The canonical choice is the Hartle-Hawking (or Euclidean) state defined by analytic continuation from Euclidean signature.  Correlation functions in this state are obtained by Wick rotation.  We start by defining
\be
t\to t_E = i t~,
\ee
to obtain the Euclidean metric
\be\label{emetric}
{ds^2_E \over \ell^2}= dr^2  +\cos^2 r dt_E^2 + \sin^2 r d\phi^2 ~.
\ee
In order for this geometry to be non-singular at $r=\pi/2$ we see that $t_E$ must be periodically identified, so that
\be\label{sdef}
(t_E,\phi) \sim (t_E,\phi) + 2\pi (m,n) ~~~~~ \forall n,m \in \Z~.
\ee With these identifications we recognize (\ref{emetric}) as the metric on $S^3$ written in Hopf coordinates.

We can then compute field theory correlation functions on the sphere $S^3$ and analytically continue them back to Lorentzian signature.  This gives field theory expectation values in a particular quantum state, which is usually referred to as the Hartle-Hawking or Euclidean vacuum state.  The physics of this state is easy to understand.  This identification \eqref{sdef} is generated by the operator 
\be
\rho=e^{-\beta H},~~~~~\beta={2\pi}
\ee
where $H$ is the Hamiltonian operator \eqref{hj}.  This is the density matrix of a canonical ensemble at fixed temperature.  Thus field theory expectation values computed in the static patch are precisely thermal.  This is the famous statement that de Sitter space is thermal; our observer at the origin $r=0$ will see the cosmological horizon emit a finite temperature bath of particles at temperature $\beta=2\pi$.

Our basic observation that there are other identifications of the $(t_E, \phi)$ coordinates which make the Euclidean geometry \eqref{emetric} smooth.  In particular, for any pair of relatively prime integers $(p,q)$ we may identify
\be\label{ldef}
(t_E,\phi) \sim (t_E,\phi) + 2\pi \left({m\over p},m{q\over p} + n\right) ~~~~~ \forall n,m \in \Z~.
\ee
These identifications define the lens space $L(p,q)$.  Comparing (\ref{ldef}) with (\ref{sdef}) we see that the lens space is the quotient $S^3/\Z_p$ of the sphere by the cyclic group of order $p$.  $L(1,0)$ is the original $S^3$.  The parameter $q$ labels different ways of embedding this cyclic group into the isometry group $SO(4)$ of $S^3$.  We note first that a shift of $q$ by a multiple of $p$ can be absorbed into a change of the parameters $n,m$.  Thus $q$ is defined only mod $p$.  Moreover, the condition that $(p,q)=1$ is necessary for the geometry to be smooth; if $(p,q)\ne 1$ one can find a pair of integers $n,m$ such that $mq = -pn$ with $0<m<p$.  This would imply that on a surface of constant $\phi$, $t_E$ is periodically identified with period less than $2\pi$, leading to a conical singularity.\footnote{More generally, for $p,q\in \R$  the interpretation is a point particle carrying mass and angular momentum. Even though the solution has a conical singularity, it is known as the Kerr-dS$_3$ \cite{Deser:1983nh}.}

From the point of view of quantum field theory the $L(p,q)$ identifications described above have a simple physical interpretation.  This can be seen by noting that the identification \eqref{ldef}
is generated by the operator
\be\label{rhois}
\rho=e^{-2 \pi \left( {1\over p}H + i {q\over p} J \right)}~.
\ee
So, just as $S^3$ defines a canonical ensemble with $\beta=2\pi$, the $L(p,q)$ define a grand canonical ensemble with temperature and angular potential 
\be\label{gc}
\beta = {2\pi\over p}~,~~~~~~\theta = 2\pi i{q\over p}~.
\ee
We note the appearance of a factor of $i$ in equation \eqref{rhois}.  This is due to the fact that angular momentum $J$ picks up a factor of $i$ on rotation to Euclidean signature; a similar factor of $i$ appears, for example, when defining the Euclidean continuation of the Kerr black hole.

%

It is important to note that we have been careful to write $L(p,q)$ as a Euclidean continuation of the static patch of de Sitter space and to make no reference to global dS${}_3$.  This is not an accident; for generic values of $p$ and $q$, $L(p,q)$ can not be described as the Wick rotation of smooth global Lorentzian geometry. 
Our point of view is that only the static patch is relevant for the study of the physics of a timelike observer.   The $L(p,q)$ should be viewed as instructions for the preparation of a state of a static patch observer.\footnote{An exception is the case $L(2,1)\sim \RP^3$, which has an interesting interpretation in terms of global dS${}_3$.  As described in \cite{Parikh:2002py}, the global Lorentzian continuation is "Schrodinger's de Sitter space", the quotient of dS${}_3$ by the antipodal map.}

Indeed, from the point of view of the static patch observer the lens spaces $L(p,q)$ have just as much a right to be called the ``Euclidean continuation of de Sitter space" as does the three sphere $S^3$.  However, correlation functions which are obtained by analytic continuation from a lens space will be different from those obtained by analytic continuation from the sphere.  So we have an apparent embarrassment of riches; of all the possible Euclidean geometries one can use to compute correlation functions, which one should we use?  

To answer this question, let us remember that the Hartle-Hawking state is defined by a path integral in Euclidean signature.  In the limit where gravity is neglected, this means that we simply Wick rotate Euclidean correlation functions computed on a fixed background geometry.  But once gravity is included the metric will fluctuate.  In a saddle point approximation, we must include contributions from all solutions to the equations of motion, and in particular all lens spaces.  So once gravity is included it is not a question of which lens space should be used.  They {\it all} contribute to the Hartle-Hawking state.  To compute correlation functions correctly we must sum over this infinite class of geometries.  This sum is the subject of the remainder of this paper.  

It is worth asking how we are to interpret this conclusion in light of the typical claim that physics de Sitter space is thermal.  As reviewed above, this is a consequence of analytic continuation from the Euclidean saddle $S^3$.  In the Euclidean sum over geometries each saddle should be weighted by (minus) its action.  This action is proportional to the volume of the saddle.  The lens spaces all have lower volume than $S^3$.  So in the limit where the de Sitter radius is large in Planck units ($\ell/G\gg1$), the contribution from the $S^3$ saddle will dominate the sum.  The contributions from the lens spaces will be suppressed by factors which are exponentially large in the volume of the sphere, which is proportional to the de Sitter entropy.
So the effects from the lens spaces are truly quantum gravitational effects which are invisible in the semi-classical limit where $G\to 0$.

\subsection{The Lens Space Farey Tail}

Before discussing details, however, we study briefly the geometric interpretation of this sum over lens spaces. 


\begin{figure}
\centering
\includegraphics[width=0.6\textwidth]{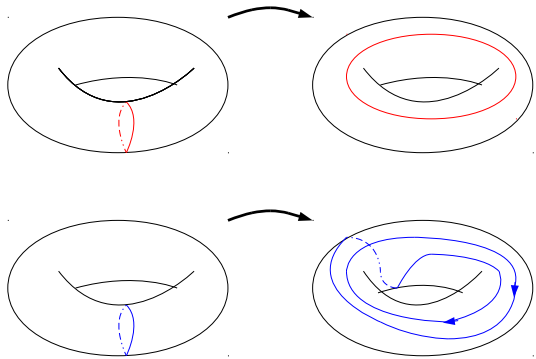}
  \caption{Lens spaces are viewed as pairs of solid tori glued together along their $T^2$ boundaries using an element of the torus mapping class group $SL(2,\Z)$.  Top: the gluing which takes the contractible cycle of one torus (red) into the dual cycle of the other torus gives the sphere $S^3$.  Bottom: a more complicated map gives a non-trivial lens space.}
\end{figure}

Topologically, a lens space can be regarded as two solid tori glued together along their $T^2$ boundaries (see figure 2).  
The different lens spaces $L(p,q)$ correspond to different ways of gluing these boundary tori together. For example, gluing the boundary tori together in the obvious way using the identity mapping gives $L(0,1)=S^1\times S^2$.  If we use the mapping which takes the contractible cycle in one torus to the dual non-contractible cycle in the other torus we obtain the sphere $L(1,0)=S^3$.  More complicated gluings correspond to more complicated choices of map used to glue together the boundary tori.

This gives a simple group theoretic classification of lens spaces.  
When we glue together the two boundary tori we must chose an element of the torus mapping class group, which is the space of smooth maps from $T^2$ into $T^2$ modulo those which are connected to the identity.  This mapping class group is $SL(2,\Z)$.  If we denote by $a$ and $b$ a basis of cycles in $H^1(T^2,\Z)$ then each element of the mapping class group gives a map 
\be
\left({a\atop b}\right) \to \left({\phantom{-}r~~s~\atop -p~~q~}\right)\left({a\atop b}\right)
\ee
for some $\left({r~s\atop -p~q}\right)\in SL(2,\Z)$.  One might therefore conclude that there is a different lens space for each element of $SL(2,\Z)$.  This is not quite the case, because on each solid torus one of the cycles in contractible; we will take this to be the $b$ cycle.  This means that the change of basis
\be
\left({a\atop b}\right) \to T \left({a\atop b}\right)~~~~~T=\left({1~~1\atop 0~~1}\right)
\ee  
leaves the topology unchanged.  We can make this change of basis for either solid torus.  So any two elements $\gamma_1,\gamma_2\in SL(2,\Z)$ such that  $T^n \gamma_1 T^m = \gamma_2$ for some $n,m\in \Z$ will lead to the same lens space. We conclude that the lens spaces $L(p,q)$ are uniquely labelled by elements of the double coset $\Z \backslash SL(2,\Z)/\Z$, where the quotient is by multiplication by $T$ on the left or the right.\footnote{Note that an element of the double coset $\Z \backslash SL(2,\Z)/\Z$ is uniquely labelled by a pair of coprime integers $(p,q)$ where $q=1,\dots,p-1$.  To see this, note that for an $SL(2,\Z)$ matrix $  \left({\phantom{-}r~s\atop -p~q}\right)$ the condition $rq+ps=1$ can be used to fix $s$ in terms of $r$, $p$ and $q$.  Moreover, this condition implies that $q$ and $p$ are coprime and that $r$ is the inverse of $q$ mod $p$.  The left quotient by $T$ identifies $r\sim r+p$, and the right quotient identifies $q\sim q+p$.  So as an element of the coset, $r$ is fixed uniquely and $q$ is defined only mod $p$.}   

This gluing picture is simply related to the explicit construction of the lens spaces given in equations (\ref{emetric}) and (\ref{ldef}).  We can divide the geometry \eqref{emetric} into two regions, one with $r<R$ and the other with $r>R$, where $R$ is between $0$ and $\pi/2$.  The region with $r<R$ is a neighborhood of the observer at $r=0$ and the region with $r>R$ is a neighborhood of the Euclidean horizon at $r=\pi/2$.  Each region is a solid torus, whose boundary is the $T^2$ of fixed radius $r=R$.   In the neighborhood of the observer the $\phi$ cycle is contractible.  In the neighborhood of the horizon the $ q t_E - p \phi$ cycle is contractible.  The full lens space geometry is found by gluing these two  regions together, giving the topological picture of Figure 2.

This is very similar to the corresponding story in the AdS case, which is known as the black hole Farey tail \cite{Dijkgraaf:2000fq}.  In that case the goal was to compute the partition function of AdS gravity by performing a sum over Euclidean geometries which are asymptotically $T^2$ at conformal infinity.  The saddle point geometries are 3-manifolds of constant negative curvature which are topologically a solid torus.  On each saddle point geometry one of the cycles of the boundary $T^2$ is contractible, and the full sum over geometries can be interpreted as a sum over all possible cycles in the boundary $T^2$.  This is a sum over a coset of $SL(2,\Z)$, the mapping class group of the boundary torus. Further, the sum over the double coset $\Z \backslash SL(2,\Z)/\Z$  in AdS gravity can be interpreted as the Farey tail expansion for an infinite family of extremal black holes \cite{Murthy:2009dq}.  
 
 We have uncovered a similar structure in de Sitter space, where the partition function is computed by summing over lens spaces.  This is regarded as the sum over possible cycles which are contractible at the Euclidean horizon.  In the analogy with the black hole Farey tail the Euclidean horizon plays the role of the ``interior" of the geometry, and the neighborhood of the observer $r<R$ plays the role of the asymptotic boundary.  
 
 We conclude with a few comments on lens space geometry which will be useful later.
The lens space $L(p,q)$ is defined by the identifications (\ref{ldef}).  To better understand these identifications we introduce the complex coordinates
\be\label{bca}
z_1 = \cos r\, e^{it}~,~~~~~z_2=\sin r\,  e^{i \phi}~.
\ee
The three-sphere is the set of points $\{|z_1|^2 + |z_2|^2 = 1\}\subset\C^2$.  These coordinates make it clear that $S^3$ can be written as a fibration of $S^1$ over $S^2$ in many different ways, by taking as our $S^1$ fibre any linear combination of the $t$ and $\phi$ directions.  The lens space identifications are 
\be\label{bcb}
\left(z_1 \atop z_2 \right)\sim\left(\omega~~0\atop 0~~\omega^q\right)\left(z_1\atop z_2\right)~,~~~~~~\omega=e^{2\pi i/p}~.
\ee 
This identifies points on the  three sphere which are related by a translation along an $S^1$ fibre by an amount $2\pi/p$; the choice of fibre is labelled by $q$.

A second simple description of lens spaces uses the fact that $S^3$ is the $SU(2)$ group manifold.  In terms of the $z_i$ coordinates defined above   a point on $S^3$ can be identified with the element 
\be
g=\left({z_1~~ z_2 \atop -{\bar z}_2 ~~{\bar z}_1}\right)\in SU(2)~.
\ee
The isometry group is then $SO(4)=SU(2)\times SU(2)/\Z_2$, which acts as 
\be
(L,R):g\to L ~g~ R~,~~~~~(L,R)\in SU(2)\times SU(2)/ \Z_2~.
\ee
The $\Z_2$ quotient arises because the element $(-1,-1)\in SU(2)\times SU(2)$ acts trivially.  The identification \eqref{ldef} which defines the lens space quotient is 
\be\label{lris}
g \sim L g R,~~~~~L=\left({\omega^{(1+q)/2}~~~~~0\atop 0~~~~~\omega^{-(1+q)/2}}\right)~,~~R=\left({\omega^{(1-q)/2}~~~~~0\atop 0~~~~~\omega^{-(1-q)/2}}\right)~.
\ee
This generates a $\Z_p$ quotient since $(L,R)\in SU(2)\times SU(2)/\Z_2$ is a $p^{th}$ root of unity.

We note that from this description it is easy to describe the isometries of a lens space.  They are those elements of $SU(2) \times SU(2)$ which commute with the left and right matrices in equation \eqref{lris}. When $q\ne \pm 1$ mod $p$ this is
$U(1)\times U(1)$. When  $q=\pm1$ mod $p$ one of the matrices in \eqref{lris} is trivial and the isometry group is $U(1)\times SU(2)$. 



\section{Partition Function: Tree Level Results}

Our goal is to evaluate the Euclidean quantum gravity path integral 
\be
Z=\int {\cal D} g \, e^{-S[g]}~=\sum_{g_c}e^{-k S^{0}+S^{1} +{1\over k} S^2 + \dots}~,
\ee
by classifying all classical saddles and computing the infinite series of perturbative corrections.  
In this section we will classify the classical saddles $g_c$ and compute the classical contribution $S^0$ to the path integral for the 3-sphere and lens spaces.   We consider here only contributions from lens spaces, whose inclusion in the path integral was motivated on physical grounds in the previous section.  It is easy enough, however, to compute the tree level contribution from all solutions; this is described in appendix A.    

We start by describing the classical saddles and computing the resulting partition sum in sections \ref{sec:saddles} and \ref{sec:sumtree}.  In section \ref{sec:CStree} we compare the tree level action to that obtained using the Chern-Simons formulation.  

\subsection{The Classical Saddles}\label{sec:saddles}

We start by considering gravity in Euclidean signature with a positive cosmological constant.  The action is
\be\label{EH}
S =- {1\over 16 \pi G} \int_{\M} d^3 x \sqrt{g} \left(R - {2\over \ell^2}\right)~.
\ee
We will use units where $\ell=1$ and write everything in terms of the dimensionless coupling $k=\ell/4 G$.  The equations of motion are 
\be\label{ca}
R_{\mu\nu} = 2 g_{\mu\nu}~.
\ee
The solutions  to \eqref{ca} are three dimensional manifolds $\M$ which are locally isometric to the three sphere $S^3$.  These geometries have been classified in the literature. We review a few relevant results here and refer the reader to e.g. \cite{thurston97} for details.  

The smooth solutions are quotients of the three sphere of the form $S^{3}/\Gamma$ where $\Gamma$ is a discrete, freely acting subgroup of the isometry group $SO(4)$ of the sphere.\footnote{In principle we could also include saddles with orbifold-type singularities coming from quotients which do not act freely.  In the absence of evidence that these geometries should be included in the path integral we will not include them, but they may be worth further study.}  These geometries are usually referred to as elliptic three-manifolds.   There are an infinite and countable number of choices for the group $\Gamma$.  In particular, $\Gamma$ must be either a cyclic group or a central extension of a dihedral, tetrahedral, octahedral, or icosahedral group by a cyclic group of even order.  This completely characterizes all possible smooth solutions to the equations of motion. 

The on-shell action of one of these saddles is proportional to its volume
\be
S[g_{c}] = -{k\over \pi} {\rm Vol}(\M)~.
\ee 
Hence for $S^{3}/\Gamma$ we have
\be\label{c:tree}
S[g_{c}] =  -{k\over \pi }{{\rm Vol}(S^3)\over |\Gamma|}=-{2\pi k\over |\Gamma|}~,
\ee 
where $|\Gamma|$ is the order of the group.  

\subsection{The Sum Over Geometries}\label{sec:sumtree}

The  contribution to the path integral of these saddles is, at tree level, equal to
\be\label{cb}
Z^{(0)}=\sum_{g_c} \,e^{-S^0} = \sum_\Gamma \exp\left({2 \pi k \over |\Gamma|}\right)~.
\ee
We note that the three sphere $S^3$ gives the dominant contribution to the partition function.  We now describe the sum over the lens spaces, which are the saddles where $\Gamma$ is abelian. In appendix \ref{quo} we describe the inclusion of the saddles with non-abelian quotients. 

For the lens space $L(p,q)$ the group $\Gamma=\Z_p$ is cyclic.  Here $p\ge 1$ is a positive integer and $q$ is a number between $1$ and $p$ which is coprime to $p$.  So the sum is
\bea\label{cc}
Z_{\rm lens}^{(0)}&=& \sum_{p=1}^\infty \sum_{q=1\atop(q,p)=1}^pe^{2\pi k/p}= \sum_{p=1}^\infty e^{2\pi k/p}\phi(p) ~.
\eea
Here $\phi(p)$ is Euler's totient function, which counts the number of integers less than $p$ and coprime to $p$. The sum is divergent, since as $p$ goes to infinity the exponential approaches one.  So this sum is dominated by terms with large $p$, i.e. by geometries whose volume is small in Planck units. 
To better understand the nature of the divergence, we will rewrite \eqref{cc} in terms of zeta functions.  Expanding the exponential we get
\bea\label{cca}
Z_{\rm lens}^{(0)}&=& \sum_{r=0}^\infty {(2\pi k)^r\over r!} \sum_{p=1}^\infty \phi(p) p^{-r}\nonumber \\
&=& \sum_{r=0}^\infty {(2\pi k)^r \over r!} {\zeta(r-1)\over \zeta(r)}~,
\eea
where we used the Dirichlet series \eqref{app:euler} and $\zeta(s)$ is the Riemann zeta function.

We can now attempt to regularize the sum (\ref{cca}).  The most natural way to do so is by using zeta function regularization, which amounts to using in (\ref{cca}) the values of the zeta function $\zeta(s)$ obtained by analytic continuation of the argument $s$.  This makes $\zeta(s)$ finite for all $s\ne 1$.
However, the zeta function has a pole at $1$ which remains even after analytic continuation.   Thus the $r=2$ term in the sum is divergent and the sum cannot be regularized using zeta function techniques. This divergence gets even worse if we add the remaining elliptic manifolds (see appendix \ref{quo:final}).
 
It is important to emphasize that, at this point, the divergence in the sum over $p$ should not worry us too much.  We have included only the tree level action, and it is natural to expect that the quantum corrections to the effective action will introduce additional $p$-dependence which might make the series more convergent.  We will see in the next section that this is precisely the case.

However, we note that  there is a clear difference with the corresponding story in AdS/CFT.  In that case a similar divergence arises from the sum over saddles which have (regularized) volume which is small in Planck units \cite{Dijkgraaf:2000fq}.  Even at tree level, however, this divergence can be removed by regulating the sum in one of a variety of ways.  It can be regulated by using the ``Farey-Tail transform" of \cite{Dijkgraaf:2000fq}, using zeta function regularization following \cite{Maloney:2007ud} or by carefully summing over the terms in the sum in the correct order \cite{Manschot:2007ha}.   In each case the answer was the same and had a natural physical interpretation.  We see that, at least for the tree level computation, the partition function of de Sitter quantum gravity can not be so regulated.

 Finally, we note one subtlety in computing the sum over geometries.  In the sum (\ref{cb}) we should in principle sum only over those geometries which are not diffeomorphic to one another, otherwise we are in danger of over counting geometries.  In three dimensions, the task of determining  which manifolds are diffeomorphic is relatively easy: manifolds are diffeomorphic if and only if they are homeomorphic.  So we should sum only over over topologically distinct manifolds.  The three manifold $S^3/\Gamma$ has fundamental group $\pi_1(S^3/\Gamma) = \Gamma$, so $S^3/\Gamma$ and $S^3/\Gamma'$ can be diffeomorphic only if $\Gamma=\Gamma'$.  So we need to ask whether it is possible for $L(p,q_{1})$ and $L(p,q_{2})$ to be diffeomorphic when $q_{1}\ne q_{2}~{\rm mod~} p$.  It turns out that these lens spaces are diffeomorphic if and only if
\be\label{qident}
q_{1}=\pm q_{2} ~{\rm mod}~p~,~~~~~{\rm or}~~~~~q_{1}q_{2}=\pm 1~{\rm mod}~p~.
\ee
It is  easy to find the diffemorphisms relating these values of $q_1$ and $q_2$ using the explicit presentation of the metric (\ref{emetric}).  For example, the diffeomorphism $\phi\to -\phi$ leads to the identification of $q$ with $-q$.  It is more challenging to show that these are the {\it only} possible diffeomorphisms; we refer the reader to \cite{thurston97} for details.

In computing the sum over lens spaces we should, strictly speaking, sum only over lens spaces modulo the relation \eqref{qident}.  In computing the exact numerical value of the partition function this is an important subtlety, but it is one that can be ignored in the present section.  The reason is that the identifications \eqref{qident} will lead to at most a factor of four in the partition sum, and will therefore appear that at the same order as the one loop results discussed in the next section.  In our present discussion of tree level results, we can therefore treat as distinct the lens spaces $L(p,q)$ for all values of $q$ mod $p$.\footnote{In fact, there is a sense in which this is the most reasonable thing to do.  The identifications \eqref{qident} are due to large diffeomorphisms which are not continuously connected to the identity.  In defining our path integral over the space of metrics we should clearly mod out by the set of local diffeomorphisms, but in principle we could simply define our set of symmetries to not include the large diffeomorphisms which lead to the identifications \eqref{qident}.  This is exactly what we do when we define the path integral of gravity in AdS/CFT; two metrics define the same state only if they are related by a diffeomorphism which vanishes sufficiently quickly at infinity.  The large diffeomorphisms which act on the boundary via conformal transformations change the state of the theory and give distinct contributions to the Euclidean path integral.   This leads to, for example, boundary gravitons (for infinitesimal conformal transformations) and the $SL(2,\Z)$ family of black holes (for large conformal transformations).  It would be interesting to define precisely the relevant symmetry group for de Sitter gravity, but we will not attempt to do so here.}
In section 4 when we compute one loop effects we will return to this correction.

\subsection{Chern-Simons Formulation}\label{sec:CStree}

It is instructive to compare the results derived above to those found using the Chern-Simons formulation of three dimensional gravity.
This is a straightforward computation at tree level, but we will go through the details explicitly in order to set the stage for less trivial uses of Chern-Simons theory in later sections.  In this section we work entirely in Euclidean signature.

The Chern-Simons formulation of three dimensional gravity is a first order formalism, where the basic variable is taken not to be the metric $g_{\mu\nu}$ but rather the frame fields $e_\mu{}^a$ and the connection $\omega_\mu{}^{bc}$, where $a,b,\dots$ are local flat indices.  These fields are typically packaged into the one forms $e^a$ and $\omega^a$ defined by:
\be
e^a = e_{\mu}{}^a dx^\mu~ ,~~~~~~ \omega^a = \frac{1}{2}\: \epsilon^a{}_{bc} \:\omega_\mu{}^{bc}dx^\mu~.
\ee
The frame fields are related to the metric in the usual way
\be
g_{\mu\nu} = e^a{}_\mu e^b{}_\nu \delta_{ab}~,
\ee
and the connection is determined by the flatness condition
\be\label{de}
de^a -  \e^a{}_{bc} \omega^b \wedge e^c=0~ .
\ee
Einstein's equation is
\be\label{do}
d\omega^a -\frac{1}{2} \e^a{}_{bc} (\omega^b\wedge \omega^c + e^b\wedge e^c) = 0~,
\ee
and the action \eqref{EH} is
\be\label{EHCS}
S=- {k\over 2 \pi} \int_\M \left(e_a\wedge (d\omega^a -\frac{1}{2}\:\e^a{}_{bc} (\omega^b\wedge \omega^c +\frac{1}{3}  e^b\wedge\: e^c )\right)~.
\ee

The remarkable observation of \cite{Achucarro:1987vz,Witten:1988hc,Achucarro:1989gm} is that equations \eqref{de}, \eqref{do} and \eqref{EHCS} are the action and equations of motion of a Chern-Simons theory.
To see this, we define the linear combinations
\be
A^a_\pm = \omega^a \pm e^a ~,
\ee
which are regarded as a pair of $SU(2)$ gauge fields.  If we introduce $SU(2)$ algebra generators $T_a$ and write
\be
A_\pm = A^a_\pm T_a~,
\ee
then the equations of motion simply become the flatness conditions for a pair of $SU(2)$ connections
\be\label{flat}
F_\pm = dA_\pm + A_\pm \wedge A_\pm = 0~.
\ee
This is the equation of motion of a Chern-Simons gauge field in three dimensions.  The action of 
such a gauge field is the Chern-Simons invariant
\be\label{csaction}
I[A] =  \int_{\M} {\rm Tr}\left( A \wedge dA + {2\over 3} A\wedge A\wedge A \right)~,
\ee
where ${\rm Tr}$ is the usual trace on the $SU(2)$ Lie algebra.  The action \eqref{EHCS} is just
\be\label{CSEH}
S =- {k\over 4\pi} (I[A_+] - I[A_-])~.
\ee
Thus Euclidean gravity is, at the level of the classical action and equations of motion, equivalent to a pair of $SU(2)$ Chern-Simons theories.

It is important to note that in Euclidean signature the Chern-Simons action is typically defined with an additional factor of $i$, so that $S_{CS}=-i {k_{\rm cs}\over 4\pi} I[A]$ where $k_{\rm cs}$ is the level of the theory.  The real part of this level must be an integer in order to insure invariance under large gauge transformations.  Euclidean gravity is related to Chern-Simons theory with purely imaginary levels \be
ik_+=k,~~~~~ik_-=-k~.
\ee  We will not attempt to study Chern-Simons theory with imaginary level non-perturbatively (see however \cite{Witten:2010cx}). Our goal is simply to use Chern-Simons formulation to compute perturbative corrections in a systematic manner.

It is illustrative to work out explicitly the Chern-Simons connections $A_\pm$ for the $S^3$ and lens space geometries.  The metric in Hopf coordinates is
\be
ds^2 = dr^2 + \cos^2 r dt_E^2 + \sin^2 r d\phi^2 ~.
\ee
In these coordinates the two circles $\theta_\pm = \phi\pm t_E$ have constant length $S^1$ fibers. 
The frame field is
\be
e = e^a T_a =   T_1\, dr+ \cos r\, T_2\, dt+ \sin r\, T_3 \,d\phi~,
\ee
and the connection is
\be 
\omega = \omega^a T_a = \cos r\, T_2\, d\phi+ \sin r\, T_3 \,dt~,
\ee
so that 
\be\label{Ais}
A_\pm = A_\pm^a T_a=\pm T_1\, dr + ( \cos r \,T_2\pm \sin r\, T_3) d\theta_{\pm}~.
\ee
It is straightforward to check that this connection is flat.  

Equation (\ref{Ais}) gives the connection both on the three sphere $S^3$ as well as on its quotients $L(p,q)=S^3/\Z_p$.
 In terms of the Hopf circles $\theta_\pm$, the lens space is defined by the  identifications
 (c.f. \eqref{bcb})
\be
\theta_\pm \sim \theta_\pm + 2 \pi {n (q\pm1) - m p\over p}~, ~~~~~\forall n,m\in \Z~.
\ee
The $m$ identification is just the usual $\phi\sim\phi+2\pi$.  The $n$ identification is the non-trivial $\Z_p$ quotient of $S^3$.


It is useful to describe the flat connection (\ref{Ais}) a bit more geometrically.  The lens space $L(p,q)=S^3/\Z_p$ has a topologically nontrivial cycle coming from the quotient by $\Z_p$.  A flat connection on $L(p,q)$ is characterized by its holonomy around this nontrivial cycle.  More precisely, the fundamental group of $L(p,q)$ is $\pi_1(L(p,q)) = \Z_p$.  An $SU(2)$ connection on $L(p,q)$ then defines a map from $\pi_1(L(p,q)) =\Z_p\to SU(2)$, defined by the holonomy of the connection around each cycle.  Since $\Z_p$ is a cyclic group, this map must take each non-trivial cycle into a $p^{th}$ root of unity in $SU(2)$.  So the image of the non-contractible cycle in $L(p,q)$ must be conjugate to a rotation by an angle ${2 \pi n\over p}$ in $SU(2)$ for some integer $n$ which is defined modulo $p$. Thus the round metric on the lens space $L(p,q)$ is characterized by a pair of integers $(n_+,n_-)$ which give the holonomy around the non-contractible cycle of the two $SU(2)$ connections.

To compute these holonomies
we note that the integral of $d\theta_\pm$ around the topologically non-trivial cycle with $(n,m)=(1,0)$ is
\be
\oint d\theta_\pm = {2\pi}{q\pm 1\over p}  ~,
\ee
so that
\be
\oint A_\pm = {2\pi}{ q\pm 1\over p} (\cos r T_2 \pm \sin r T_3)~.
\ee
The holonomy of the gauge field is
\be
\exp{\oint A_\pm} = \cos\left( {2\pi}{q\pm 1\over p}\right) + (\cos r T_2 \pm \sin r T_3) \sin\left( 2\pi{q\pm 1\over p}\right)~.
\ee
In writing this we have used the formula
\be
e^{u \theta} = \cos \theta + u \sin \theta~,
\ee
for any $u = u^a T_a$ such that $u^a u_a =1$.  
We see, as expected, that $e^{\oint A_\pm}$ is a root of unity in $SU(2)$.
We conclude that the lens space $L(p,q)$ corresponds to a pair of $SU(2)$ gauge fields $A_\pm$ with holonomy\footnote{
We note that these are half-integer rather than integer because $SO(4)$ is actually the $\Z_2$ quotient $SO(4)=SU(2)\times SU(2)/\Z_2$.  So the holonomy of the connection is $p^{th}$ root of unity in $SO(4)$ even though $n_L$ and $n_R$ are in some cases half-integer. }
\be\label{hol}
(n_+,n_-) = \left({q + 1 \over 2},{q-1\over 2}\right)~.
\ee
In fact, we could have concluded this without doing any work.  If we regard the three sphere as the $SU(2)$ group manifold, the two $SU(2)$ Chern-Simons connections are associated with the group actions by left and right multiplication.  The holonomies can then be read off from \eqref{lris}.

We can now go ahead and compute the action of the Chern-Simons theory with this connection.
The Chern-Simons invariant of an $SU(2)$ gauge field on a lens space $L(p,q)$ with holonomy $n$ is \cite{Jeffrey:1992tk}
\be\label{inv}
 {1\over 8\pi^2}\int {\rm Tr}\left( A \wedge dA + {2\over 3} A\wedge A\wedge A  \right)= {q^* \over p}n^2~,
\ee
where $q^*$ is the inverse of $q$ mod $p$:
\be
q^*q=1\,{\rm mod}\,p~.
\ee
Plugging this into the Chern-Simons action \eqref{EHCS} we see that this 
reproduces the correct gravity action  \eqref{c:tree}
\bea
Z_{(p,q)}^{(0)} &= &\exp\left (-S[g_{(0)}]\right)= \exp\left(i{k_+\over 4 \pi}I[A_+] + i{k_-\over 4 \pi} I[A_-]\right)\cr&=&
\exp\left(\pi i {q^*\over 2 p} \left({k_+(q+1)^2 +k_-(q-1)^2} \right)\right)\cr &= &\exp\left({2\pi k \over p}\right)~.
\eea

\section{Partition Function: One-loop Results}

We turn now to the evaluation of quantum corrections to the partition function at the one-loop level, i.e. the computation of $S^1$ in (\ref{Zsaddle}). We will start by computing the answer directly in gravity, evaluating the appropriate one-loop determinants using heat kernel techniques.  In section \ref{sec:sumloop} we compute the sum over geometries including this one-loop contribution.  In section \ref{sec:CSloop} we check this answer by comparing with the results in Chern-Simons theory.

\subsection{Gravity Computation}\label{heat}

\subsubsection{\it One-loop determinants in Einstein gravity}

The one loop partition function of Einstein gravity has been considered by various authors; see \cite{Gibbons:1978ji,Christensen:1979iy,Yasuda:1983hk} for discussion of the D-dimensional case.  We only summarize a few results here.

The one-loop contribution $S^1$ to the path integral
\be
\int {{\cal D} g\over V_{\rm diff}} e^{- S}= \sum e^{-k S^0+S^1+\dots}
\ee
is obtained by integrating over the linearized fluctuations around each classical saddle. The measure factor $V_{\rm diff}$ reflects the fact that we integrate only over orbits of the diffeomorphism group in the space of metrics.  At the linearized level a diffeomorphism generated by the vector $V_\mu$ takes 
\be\label{diff}
g_{\mu\nu}~\to~g_{\mu\nu} + \nabla_{(\mu} V_{\nu)}~.
\ee
This can be used to impose a gauge condition on the linearized metric fluctuations.  A standard choice is transverse gauge.  Linearizing the action and computing the appropriate gaussian integrals in this gauge we obtain a ratio of functional determinants
\bea\label{tvsdet}
Z^{(1)}=e^{S^{(1)}}={\det\phantom{\,}\left(\Delta_{(1)}^{LL}-{2\over 3}R\right)\over \det\phantom{\,}^{1/2}\left(\Delta_{(2)}^{LL}-{2\over 3}R\right)\det\phantom{\,}^{1/2}\left(\Delta_{(0)}^{LL}-{2\over 3}R\right)}~.
\eea
The denominator comes from linearized metric fluctuations, which have been decomposed into a transverse traceless part and a scalar part coming from the trace of the metric. The numerator is the Fadeev-Popov determinant which arises when we gauge fix, and can be regarded as the contribution of a spin-1 ghost.

The operators in (\ref{tvsdet}) are obtained by linearizing the action and are defined as follows.  $R$ is the Ricci scalar.  The operator $\Delta^{LL}_{(2)}$ acts on symmetric, traceless 2-tensors and $\Delta^{LL}_{(1)}$ acts on both the transverse and longitudinal components of a vector.  
They are Lichnerowicz Laplacians which are written in terms of the usual Laplacian $\Delta_{(j)}=\nabla^\alpha\nabla_\alpha$ acting on a field of spin $j$ as:\footnote{We note that $\Delta^{LL}_{(p)}$ coincides with the Hodge-de Rham Laplacian when acting on $p$-forms.} 
\bea
\Delta_{(2)}^{LL}T_{\mu\nu} &=&-\Delta_{(2)} T_{\mu\nu}-2R_{\mu\alpha\nu\beta}T^{\alpha\beta}+R_{\mu\alpha}T^\alpha_{~\nu}+R_{\nu\alpha}T^\alpha_{~\mu}~, \cr 
\Delta_{(1)}^{LL}T_\mu&= &-\Delta_{(1)} T_{\mu}+R_{\mu\alpha}T^\alpha~,\cr
\Delta_{(0)}^{LL}T&= &-\Delta_{(0)} T~.
\eea

In the absence of zero or negative modes \eqref{tvsdet} can be simplified further.  This follows from the harmonic decomposition of tensors, which is reviewed in Appendix B.1.  This decomposition allows us to cancel common factors in the numerator and denominator of \eqref{tvsdet} to obtain
\bea\label{detloop}
Z^{(1)}=\sqrt{{\det'\phantom{\,}\left(\Delta_{(1)}^{LL}-2R/D\right)_{T}\over \det '\phantom{\,}\left(\Delta_{(2)}^{LL}-2R/D\right)_{TT}}}~.
\eea
Here prime denotes only the positive eigenvalues of the operators and the subscript $T$ denotes the transverse part. In three dimensions 
\bea\label{curvature}
R_{\mu\alpha\nu\beta}={R\over6}(g_{\mu\nu}g_{\alpha\beta}-g_{\mu\beta}g_{\nu\alpha})~,\quad R_{\mu\nu}={R\over3}g_{\mu\nu}~,\quad R=6~,
\eea
so the above expressions simplify to
\bea
\left(\Delta_{(2)}^{LL}-{2\over 3}R\right)T_{\mu\nu} &=(-\Delta_{(2)}+2)T_{\mu\nu}~,\cr 
\left(\Delta_{(1)}^{LL}-{2\over 3}R\right)T_\mu&= (-\Delta_{(1)}-2)T_{\mu}~.
\eea

Formula \eqref{detloop} is perfectly correct if all of the relevant operators have positive definite spectrum. However, on the compact manifolds of interest this is not quite  the case. We must include in \eqref{detloop} corrections coming from the non-positive eigenvalues of the operators
\be\label{ghost}
\Delta_{(0)}^{LL}-{2\over 3}R~,\quad \Delta_{(1)}^{LL}-{2\over 3}R~.
\ee  

Let us first consider the vector operator in \eqref{ghost}.  It is easy to show that for the spherical three manifolds under consideration this operator has no negative modes.  The zero modes of this operator are Killing vectors.\footnote{ To prove this, note that the Killing's equation $\nabla_{(\mu} V_{\nu)}= 0$ along with \eqref{curvature} give
$$
-\nabla_{\mu}\nabla^\mu V_\nu-R_{\mu\nu}V^\mu=0~.
$$
}
By construction, these zero modes are not included in the vector determinant in \eqref{tvsdet}.  Indeed, from equation \eqref{diff} we see that a Killing vector  (KV) $V_\mu$ generates the trivial diffeomorphism.  This means that our gauge fixing procedure is slightly ill-defined.  In writing \eqref{tvsdet} we have introduced gauge-fixing terms which define sections in the space of metrics which are supposed to intersect each orbit of the diffeomorphism group exactly once.  Metrics which are related by an isometry are of course diffeomorphic, but this has been missed by our gauge fixing procedure.  This can be repaired by splitting $V_{\rm diff}$ into two parts, one coming from isometries and one coming from the gauge condition:
\be\label{measure}
\int {{\cal D} g\over V_{\rm diff}} = \int {{\cal D} g\over V_{\rm KV} V_{\rm gauge}}=\int {{\cal D}h\det_{\rm ghost}\over kV_{\rm KV}}~.
\ee
Here $h$ is a linearized gauge fixed metric and $\det_{\rm ghost}$ is the vector determinant appearing in \eqref{ghost}. We also included a factor of the coupling to account for the normalization of the metric fluctuations.  This gives a correction to \eqref{detloop} for each Killing vector; we must include the volume $V_{\rm KV}$ of the isometry group.
 
We now turn to the scalar Laplacian.  For a spherical manifold the constant mode will lead to single negative eigenvalue for the scalar operator in \eqref{ghost}.  There will be additional negative modes coming from conformal Killing vectors (CKVs).  
To see this, we note that for every solution of the CKV equation
\be
\nabla_{(\mu}V_{\nu)}-{2\over D}g_{\mu\nu}\nabla_\alpha V^\alpha=0~,
\ee 
the scalar $\phi=\nabla_\alpha V^\alpha$ will be an eigenmode of the scalar Laplacian in \eqref{ghost} with negative eigenvalue. 
In both of these cases, the path integral now appears to contain a gaussian integral with the wrong sign.  This can be remedied by rotating the contour of integration in field space by 90 degrees, turning this into a convergent integral.  This is a standard procedure in gravitational path integrals, following \cite{Gibbons:1978ac} (see also \cite{Polchinski:1988ua}).  

Our final expression for the one-loop determinant is
\bea\label{HKa}
Z^{(1)}= D_{zm} \sqrt{{\det\phantom{\,}'\left(-\Delta_{(1)}-2\right)_T\over \det\phantom{\,}'\left(-\Delta_{(2)}+2\right)_{TT}}}~, ~~~~~D_{zm}={\sqrt{\det {\rm CKV}}\over k V_{KV}}
\eea
where $D_{zm}$ is the contribution from zero and negative modes.

\subsubsection{\it Heat kernels and functional determinants}

We now compute the one-loop determinants in \eqref{HKa}. 
To obtain the eigenvalues and degeneracies of the operators appearing in this equation we will use heat kernel techniques.  

For the differential operator $\Delta_{(j)}$ we define the heat kernel\footnote{The heat kernel computed in \cite{David:2009xg},  and used here, uses the Hodge-de Rham decomposition for tensors, e.g. $K^{(1)}$ is the heat kernel for a transverse vector, and $K^{(2)}$ is the kernel for a symmetric, traceless and transverse 2-tensor and so forth. }
\be\label{kdef}
K^{(j)}(x,y;t)=\langle y|e^{t\Delta_{(j)}} |x\rangle=\sum_n\psi_n^{(j)}(x)\psi^{(j)}_n(y)^{*}e^{\lambda_n^{j}t}~,
\ee
where $\psi^{(j)}_n(x)$ and $\lambda_n^j$ are the eigenfunctions and eigenvalues of $\Delta_{(j)}$. The heat kernel $K^{(j)}(x,y;t)$ obeys the heat equation, which is the statement that it is annihilated by the differential operator $\p_t - \Delta_{(j)}$.  If we integrate over space and use the orthonormality of the eigenfunctions we obtain
\be\label{kdef2}
K^{(j)}(t)\equiv \int d^3x\sqrt{-g}\,K^{(j)}(x,x;t)~ = \sum_n d_n e^{\lambda_n^{j} t}. 
\ee
This is a function of $t$ which encodes the spectrum of the operator $\Delta_{(j)}$.
The utility of this method is that heat kernel \eqref{kdef} on the sphere is relatively easy to compute, either by constructing eigenfunctions or using the description of $S^3$ as the $SU(2)$ group manifold.  The heat kernel on the lens space is then by found by using the method of images.  This was done explicitly by \cite{David:2009xg}; we refer the reader there for details.

The operators of interest all have an infinite number of eigenvalues, so the one loop determinants must be regulated carefully.
We will use zeta function regularization, following \cite{Hawking:1976ja}.
Let us consider a differential operator with eigenvalues $\lambda_n$ which have degeneracies $d_n$.  The logarithm of the functional determinant is 
\be\label{log}
\log \det = \sum_{n}d_n\ln(\lambda_n)~.
\ee
To regulate the sum over $n$ we define the zeta function
\be\label{zetahk}
\zeta(s)_{HK}=\sum_{n}{d_n\over\lambda_n^s}~,
\ee
The identity
\be\label{Z1zhk}
{d\over ds}\zeta(0)_{HK}=-\sum_{n}d_n\ln(\lambda_n)~.
\ee
%
%
%
can then be used to compute the determinant \eqref{log}.
In general, the sum \eqref{zetahk} converges only when the real part of $s$ is sufficiently large.  However, we can regard $\zeta(s)_{HK}$ as the function on the complex $s$-plane obtained by analytic continuation of the sum for large $s$.  With this definition, equation \eqref{Z1zhk} provides a regulated version of the determinant.

Comparing \eqref{kdef2} and \eqref{zetahk}, we see that the zeta function is related to the heat kernel by the integral
\be
\zeta(s)_{HK} = {1\over \Gamma(s)} \int_{0}^{\infty}t^{s-1}K^{(j)}(t)
\ee
Thus
\bea\label{HKb}
\log\left[\det(-\Delta_{(j)}+m_j^2)\right]=-{d\over ds}\left({1\over \Gamma(s)}\int_{0}^{\infty}t^{s-1}K^{(j)}(t)e^{-m_j^2 t}dt\right)_{s=0}
\eea
The one loop determinant \eqref{HKa} of three dimensional gravity is
\bea\label{Z1}
\log { Z^{(1)}}&=&-{1\over 2}\log[\det\,'\left(-\Delta_{(2)}+2\right)]+{1\over 2}\log[\det\,'\left(-\Delta_{(1)}-2\right)] \cr &=& {1\over 2}{d\over ds}\left({1\over \Gamma(s)}\int_{0}^{\infty}t^{s-1}K^{(2)}(t)e^{-2t}dt-{1\over \Gamma(s)}\int_{0}^{\infty}t^{s-1}K^{(1)}(t)e^{2t}dt\right)_{s=0}~.
\eea
For simplicity we have suppressed the factors of $V_{\rm KV}$ and $\det{\rm CKV}$ in \eqref{HKa}.
We now need explicit expressions for the heat kernels on the lens spaces $L(p,q)$.  

%
%

%

We first consider the simple case of the 3-sphere. 
The heat kernel for a bosonic field with spin $j\ge0$ is%
\bea\label{Ksphere}
K^{(j)}(t)&=&(2-\delta_{j,0})\sum_{n=j+1}^\infty (n^2-j^2)e^{E_n^jt}~,
\eea
where
\be\label{En}
E_n^j=-n^2+j+1~.
\ee

It is worth noting that from \eqref{Ksphere} we can derive explicitly the zero and negative modes discussed above.
At large $t$ the vector heat kernel becomes $K^{(1)}\sim 6 e^{-2t}$, so that $\Delta_{1}+2$  has six zero modes corresponding to the six Killing vectors of $S^3$.  Likewise, the scalar heat kernel behaves as $K^{(0)} \sim 1 + 4 e^{-3t}+\dots$; these coefficients come from the constant mode and the four CKVs of the sphere.

Using \eqref{Ksphere} in \eqref{Z1} we get
\bea
\log Z^{(1)}_{S^3} = -\sum_{n=3}^\infty\left[(n^2-4)\ln(n^2-1)-(n^2-1)\ln(n^2-4)\right]~.
\eea
The corresponding zeta function is 
\bea
\zeta(s)_{S^3}&=&-\sum_{n=3}^\infty\left({n^2-4\over(n+1)^s} +{n^2-4\over(n-1)^s} \right) +\sum_{n=3}^\infty\left({n^2-1\over(n+2)^s} +{n^2-1\over(n-2)^s} \right)\cr &=& 12\zeta(s)-{2\over 2^s}-{3\over 4^s}~.
\eea
where $\zeta(s)$ is the Riemann zeta function and we dropped terms independent of $s$. Using \eqref{zeta} and \eqref{Z1zhk}, we find  
\bea
 Z^{(1)}_{S^3}= {\pi^6\over 4}~.
\eea

For a lens space we can use this same technique to compute the regularized determinant. Defining   
\be
\tau= \tau_1-\tau_2 ~,\quad \bar\tau= \tau_1+\tau_2~,~~~~~~\tau_1={2\pi q\over p}~,\quad \tau_2={2\pi \over p}~,
\ee
the heat kernel is \cite{David:2009xg}
\bea\label{Klens}
K^{(j)}(t)={1\over p}(1-{\delta_{j,0}\over 2})\sum_{n=j+1}^\infty d_{n}^{(j)} e^{E_n^jt}~,
\eea
where $E_n^j$ is given by \eqref{En} and 
\bea\label{dnj}
d_{n}^{(j)}=\sum_{m\in \Z_p}{1\over  \sin{m\tau\over 2}\sin{m\bar\tau\over 2}}\left[\cos(jm\tau_1)\cos(nm\tau_2)-\cos(jm\tau_2)\cos(nm\tau_1)\right]~.
\eea
%
%
We note that, as above, we can extract from this expression the number of Killing vectors and conformal Killing vectors of the lens space.   The $n=2$ term in $K^{(1)}(t)$ gives the number of Killing vectors of $L(p,q)$, which is
\be
{2\over p}\sum_{m\in Z_p}(1+\cos(m\tau)+\cos(m\bar\tau))=2(1+\delta_{q,1} +\delta_{q,p-1} )~.
\ee
This agrees with the fact (noted below equation \eqref{lris}) that when $q\ne \pm 1$ mod $p$ the isometry group is $U(1)\times U(1)$ and when $q=\pm 1$ mod $p$ the isometry group is $SU(2)\times U(1)$.  Likewise, 
the CKVs are given by the $n=2$ term in $K^{(0)}(t)$.  From \eqref{Klens} this term is exactly zero; lens spaces do not have CKVs.  Indeed, one can check explicitly that all four of the CKVs of the sphere are removed by the quotient by $\Z_p$.

%
%
%
 %
 
 The one-loop determinant is
\bea\label{K1}
\log{Z^{(1)}_{\rm lens}}={1\over 2p}\sum_{n=3}\left[d_{n}^{(1)}\ln(n^2-4)-d_{n}^{(2)}\ln(n^2-1)\right]~.
\eea
We must now regulate \eqref{Z1} and construct the appropriate zeta function.  A detailed derivation of the zeta function is given in Appendix \ref{app:HK}.  The result is
\bea\label{ffzhk}
&&\zeta(s)_{\rm lens}=p^{-s}\sum_{\pm}\Bigg[\zeta(s,\pm{q^*-1\over p})+\zeta(s,\pm{q^*+1\over p})+\zeta(s,\pm{q-1\over p})+\zeta(s,\pm{q+1\over p})\cr 
&&\quad \quad\quad\quad\quad \quad\quad-\left({p\over \pm 1-q^*}\right)^s-\left({p\over \pm1- q}\right)^s\Bigg]+4p^{-s}\zeta(s)-{1\over 4^s}~,
\eea
when  $q\pm 1$ is not a multiple of $p$, and $\zeta(s,a)$ is the Hurwitz zeta function. $\zeta(s)_{\rm lens}$ includes both the $j=1,2$ contributions from the heat kernel.  
Differentiating \eqref{ffzhk} we get
\bea\label{dzhk}
{d\over ds}\zeta(0)_{\rm lens}=-\sum_{\pm}\ln\left({16\pi^2\over p^2}\sin\left(2\pi {q^*\pm1 \over p}\right)\sin\left(2\pi{ q\pm1 \over p}\right)\right)~,
\eea
where we have used \eqref{hurwitz} and \eqref{gg}.
Exponentiating \eqref{dzhk} we get
\be\label{Z1lensa}
Z^{(1)}_{\rm lens}={4\pi^2\over p^2}\left[\cos\left({2\pi\over  p}\right)-
\cos\left({2\pi q\over  p}\right)\right]
\left[\cos\left({2\pi\over  p}\right)-
\cos\left({2\pi q^*\over  p}\right)\right]~.
\ee
This is the one-loop determinant for $L(p,q)$, valid for $p>2$ and $q \neq \pm 1 \mod p$. 

 If we take $q=\pm 1 \mod p$ in \eqref{Z1lens} several of the steps used to obtain \eqref{ffzhk} break down. The correct zeta function in this case is
 given by \eqref{zhkq1} 
\bea\label{zhkq}
\zeta(s)_{(p,1)}&=&\zeta(s)_{(p,p-1)}\cr
&=&2p^{-s}\left[\zeta(s,{2\over p})+\zeta(s,-{2\over p})-\left(-{p\over 2}\right)^s\right]+8p^{-s}\zeta(s)-{2\over 4^s}-{1\over 2^s}~,
\eea
and the one-loop contribution is
\be\label{Zq1a}
Z^{(1)}_{(p,1)}=Z^{(1)}_{(p,p-1)}={2\pi^4\over p^4}\sin^2\left({2\pi\over  p}\right)~.
\ee
when $p>2$.

The case $p=2$ and $q=1$ must be treated separately.  The zeta function is given by \eqref{zhkq2} and the partition function is
\be\label{Z21}
Z^{(1)}_{(2,1)}={\pi^6\over 2^8}~.
\ee

\subsubsection{\it Volume of zero modes}

We now need to compute the prefactor $D_{zm}$ in \eqref{HKa} which comes from the zero and negative modes.
As lens spaces do not have conformal Killing vectors we need only to compute the volume $V_{\rm KV}$ of the isometry group.
  
The Killing vectors generate the isometry groups are $U(1)\times U(1)$ or $U(1)\times SU(2)$, depending on whether or not $q=\pm1$ mod $p$.  In computing the volume of the the isometry groups we must take care to normalize our Killing vectors appropriately. 
In doing so we will follow the logic of  \cite{Gibbons:1978ji,Carlip:1992wg}. 
Each Killing vector is normalized so that the integral of the norm of the volume of the manifold is fixed.
Thus
 \be
 V_{\rm KV}= ({\rm Vol}(S^3/\Gamma))^{n_k/2}= \left({2\pi^2\over |\Gamma|}\right)^{n_k/2}~,
 \ee
 where $n_k$ is the number of Killing vectors of $S^3/\Gamma$. 
 
Incorporating this factor in \eqref{Z1lensa} and  \eqref{Zq1a}
\be\label{Z1lens}
Z^{(1)}_{\rm lens}={2\pi\over kp}\left[\cos\left({2\pi\over  p}\right)-
\cos\left({2\pi q\over  p}\right)\right]
\left[\cos\left({2\pi\over  p}\right)-
\cos\left({2\pi q^*\over  p}\right)\right]~,
\ee
 and
 \be\label{Zq1}
Z^{(1)}_{(p,1)}=Z^{(1)}_{(p,p-1)}={\pi\over 2kp^2}\sin^2\left({2\pi\over  p}\right)~,
\ee
The results for $S^3$ and $L(2,1)$ give
 \be\label{Z21a}
 Z^{(1)}_{S^3}={\pi^3\over 2^5k}~\quad Z^{(1)}_{(2,1)}={\pi^3\over 2^{11}k}~.
 \ee
  
Finally, we note that in the above discussion we have included only those isometries which are connected to the identity.  There are also discrete isometries not connected to the identity, which contribute an additional finite factor to $V_{\rm KV}$.  For a general lens space there are four such discrete symmetries; these are precisely the discrete symmetries which lead to the identifications between lens spaces described in equation \eqref{qident}.  For example, the reflection $\phi\to-\phi$ takes takes $L(p,q)$ to $L(p,p-q)$.   In our sum over geometries we have chosen to sum over all coprime values of $(p,q)$ without enforcing condition \eqref{qident}.  Thus we should in principle divide \eqref{Zq1} by an additional factor to account for this.  This will lead to at most a factor of four, and will not affect the qualitative results of our analysis.  We will therefore omit this factor in what follows.
 
 \subsection{Regulating the partition function}\label{sec:sumloop}

Gathering our results, the partition function including the tree level and one-loop contributions takes the form
\bea\label{ZS3L}
Z=Z_{S^3}+Z_{\rm lens}
\eea
with
\bea
Z_{S^3}=Z_{S^3}^{(0)}Z_{S^3}^{(1)}={\pi^3\over 2^5}e^{2\pi k}~,
\eea
and
\bea\label{Zlensfinal}
Z_{\rm lens}= \sum_{p=1}^\infty\sum_{(p,q)=1}e^{2\pi k/p} Z^{(1)}_{\rm lens} +\sum_{p=1}^\infty e^{2\pi k/p}  Z^{(1)}_{ (p,1)}+\sum_{p=1}^\infty e^{2\pi k/p} Z^{(1)}_{(p,p-1)}+e^{\pi k} Z^{(1)}_{ (2,1)}~,
\eea
where we used that the tree level contribution is \eqref{c:tree} and the one-loop terms are given by \eqref{Z1lens}, \eqref{Zq1} and \eqref{Z21a}.

As we showed in section \ref{sec:sumtree} the tree level sum over $p$ is divergent because spaces with small volume dominate the partition function. This divergence was not cured by zeta function regularization.  Now that we have included the proper measure and quantum corrections we can ask if the sum is more convergent.  We start by looking at each contribution to \eqref{Zlensfinal} separately, starting from the $q=\pm 1 \mod p$ terms which are proportional to
\be
\sum_{p=1}^\infty{e^{2\pi k/p}\over p^2}\sin^2({2\pi\over p})~.
\ee 	 
This sum is absolutely convergent,  and in particular the very ``quantum'' saddles are suppressed by $p^{-2}$.  Quantum corrections have drastically modified the convergence of the series for this class of instantons. 

The first term in \eqref{Zlensfinal}  ($q\neq \pm 1\mod p$), after summing over  $q$, is given by 
\bea\label{Ztot}
\sum_{p=1}^\infty{e^{2\pi k/ p}\over p}\Bigg[\cos^2\left({2\pi\over p}\right)\phi(p)-2\cos\left({2\pi\over p}\right)\mu(p)+{1\over 2}(S(1,1,p)+S(1,-1,p))\Bigg]~,
\eea
where $\phi(p)$ is the Euler's totient function as introduced in \eqref{cc}. $S(a,b,m)$ is the Kloosterman sum 
and $\mu(m)$ is the Mobius function which we briefly review in appendix \ref{app:formulas}.
%
We will consider each term separately. The terms proportional to the Kloosterman sum in  \eqref{Ztot} are
\bea\label{regA}
 \sum_{p=1}^\infty {1\over p}\,e^{2\pi k/p}S(1,\pm1,p)=\sum_{r=0}^\infty {\left(2\pi k\right)^r\over r!} \sum_{p=1}^\infty p^{-r-1} S(1,\pm1,p)~.
\eea
As we explained in appendix \ref{A:Kloo} the generating function for $S(1,\pm1,p)$ has no poles for positive integral values of $(r+1)$.  Thus the sum can be regulated. Similarly, the term in \eqref{Ztot} proportional to the Mobius function
\bea
 \sum_{p=1}^\infty {1\over p}\,e^{2\pi k/p}\cos\left({2\pi\over p}\right)\mu(p)=\sum_{m,n=0}^\infty (-1)^n{\left(2\pi \right)^{m+2n}\over m!(2n)!} {k^m\over \zeta(m+2n+1)}~,
\eea
can also described by an analytic function with no poles (see \eqref{app:mu}).  So this term can also be regulated.

We are left with the first term in \eqref{Ztot}, which up to an overall constant is
\bea\label{zz}
\sum_{p=1}^\infty{1\over p}\,e^{2\pi k/ p}\cos^2\left({2\pi\over p}\right)\phi(p)~.
\eea
Expanding both the exponential and cosine function we get
\bea\label{reg}
{1\over 2}\sum^\infty_{n,m=0}(-1)^n{(4\pi)^{2n}\over (2n)!}{(2\pi k)^{m}\over m!}{\zeta(m+2n)\over \zeta(m+2n+1)}+{1\over 2}\sum^\infty_{m=0}{(2\pi k)^{m}\over m!}{\zeta(m)\over \zeta(m+1)}~.
\eea
The denominators in these expressions are finite and non-zero for all values of $n$ and $m$.  However, the analytic continuation of $\zeta(s)$ has a pole at $s=1$ leading to divergences from the   $n=0$ and $m=1$ terms.

This implies that the inclusion of one-loop effects, while they make the sum over geometries more convergent, still do not allow us to regulate the partition function using standard techniques.  Explicitly, 
\be
Z= {24} \zeta(1) + \dots
\ee
where $\dots$ denote terms which are finite upon zeta function regularization.
One might hope that there might be another regularization scheme that will cure this divergence, but that does not seem feasible. Note that  the phases in \eqref{zz} are all positive,  implying that there is no obvious  re-ordering of summations involved in $Z_{\rm lens}$ that will regulate the infinity.  This is in contrast with analogous computations of the elliptic genus in the black hole Farey tail, where a delicate cancelation of phases could render the sum regularizable. 

In the following section we will demonstrate that this divergence persists even when all order loop effects are included.  We will comment more on the nature and implications of this divergence in the discussion.

\subsection{Comparison with Chern-Simons Formulation}\label{sec:CSloop}

Before proceeding to the all loop results, it is useful to check that the one-loop expressions derived above agree with those computed using the Chern-Simons formulation.  As reviewed above, the action and equations of motion of Einstein gravity are equivalent to two copies of $SU(2)$ Chern-Simons theory at levels $\pm ik$. It is worth stressing that at the non-perturbative level Chern-Simons theory and gravity  do not appear to be equivalent (for a more detailed discussion, see \cite{Witten:2007kt}).  But at the level of perturbation theory the rewriting of the metric variables in terms of the connection variables is straightforward, so we expect that the two theories should agree to all orders in perturbation theory around a given saddle.  In this section we check this agreement explicitly at the one loop level.  
  
  The advantage of the Chern-Simons approach is that it is relatively easy to compute the relevant partition functions, following \cite{Witten:1988hf}. The $SU(2)$ Chern-Simons partition function on $L(p,q)$ was computed in \cite{Jeffrey:1992tk}.  This exact answer can then be reorganized so that it looks like a sum over classical saddles, i.e. a sum over flat $SU(2)$ connections on $L(p,q)$.  Each saddle is then weighted by its classical action along with an (in principle infinite) series of perturbative corrections.  All perturbative corrections are computed in one fell swoop using the techniques of topological quantum fields theory. The only tricky part is to isolate the correct contribution which comes of the flat connection corresponding to the usual metric on the lens space. 
  
For an $SU(2)$ Chern-Simons theory on a lens space, a flat connection is labelled by an integer $n$ which gives the holonomy of the connection around the non-contractible cycle, as described in section 3.3.  In the large $k$ limit, the contribution to the partition function of one of these flat connections is  \cite{Freed:1991wd,Jeffrey:1992tk}
\bea\label{CSW}
Z_{\rm CS}&\approx&
i\sqrt{2\over k_{\pm}p}\sum_{n=1}^p\exp\left(2\pi ik_{\pm}{ q^* n^2\over p}\right)\sin\left(2\pi {q^*n \over p}\right)\sin\left(2\pi{ n \over p}\right)~.
\eea
This encapsulates the tree and one-loop expressions. Using \eqref{CSEH} and isolating the contribution of the flat connection with holonomy  \eqref{hol}, from  \eqref{CSW} the contribution of $L(p,q)$ to the gravitational partition function is
\bea\label{Z1lensCS}
Z_{(p,q)}&=&\exp\left(-kS^{(0)}+S^{(1)}\right)\cr
&=& {1\over 2kp}e^{2\pi k/ p}\left[\cos\left({2\pi\over  p}\right)-
\cos\left({2\pi q\over  p}\right)\right]
\left[\cos\left({2\pi\over  p}\right)-
\cos\left({2\pi q^*\over  p}\right)\right]~.\eea
This expression exactly agrees with the gravitational result \eqref{Z1lens}, up to numerical factors that are independent of $p$ and $q$. 

It is worth commenting on some features of the derivation of \eqref{CSW} in the Chern-Simons theory \cite{Witten:1988hf,Freed:1991wd}. In the perturbative expansion of the path integral, the one-loop contribution involves a product of determinants which turn out to be  the square root of the Ray-Singer torsion. One could wonder, as we did for the gravity calculation, if these determinants have zero modes, i.e. if there is a residual gauge symmetry that leaves the connection invariant.  A simple  computation shows that this is not the case. This implies that when going from the metric formulation gravity to the first order formalism, the ambiguities in the gauge fixing procedure of the metric due to Killing vectors disappear and there is no need to include an integral over a space of collective coordinates.

We also note that the gravitational interpretation of \eqref{Z1lensCS} when $q=\pm 1\mod p$ is a bit subtle, since in this case one of the holonomies (either $n_+$ or $n_-$) vanishes and the connection is trivial.  This does not imply that the path integral of CS is zero, but it does mean that the constant piece in large $k$ expansion is ill defined.  As we will show in the next section, the all loop invariant is non-zero for  $q=\pm 1\mod p$. 
   
\section{Partition Function: All Loop Results}

We now use the Chern-Simons formulation to compute quantum corrections to the saddle point action at all orders in perturbation theory. In prior sections we reviewed the gravity/Chern-Simons theory dictionary and checked the equivalence at tree and one-loop level.  We now apply this relation at all orders, which combined with our classification of classical saddles gives a complete computation of the gravitational path integral over lens spaces.

The exact partition function of $SU(2)$ Chern-Simons theory on a lens space is \cite{Jeffrey:1992tk}
\bea\label{fullCSW}
Z_{\rm CS}&=&\int {\cal D}A \exp\left({ik_{\rm cs}\over4\pi}I[A]\right)\cr &=&
{-i\over \sqrt{ 2rp}}\exp\left({6\pi i s(q^*,p)\over r}\right)\sum_{\pm}\sum_{n=1}^p\pm\exp\left(2\pi ir{ q^* n^2\over p}\pm{\pi i \over r p}\right)\cos\left(2\pi {(q^*\pm1)n \over p}\right)~.
\eea
The Dedekind sum $s(q,p)$ is defined in appendix C.1 and $r$ is related to the level $k_{\rm cs}$ as
\be\label{rk}
r\equiv k_{\rm cs}+2~.
\ee  
The sum over $n$ is a sum over saddle points, i.e. a sum over flat connections.  Each term in the sum represents the classical action (the Chern-Simons invariant) for this saddle along with all perturbative corrections in powers of ${k_{cs}}$. 

It is important to note that when the connection is trivial ($n=0\mod p$)  the invariant \eqref{fullCSW} is not zero as one might have concluded from \eqref{CSW}. Instead the $n=0$ saddle point contributes a factor
\be
Z_{\rm CS}(n=0)=
\sqrt{{2\over rp}}\exp\left({6\pi i s(q^*,p)\over r}\right)\sin\left({\pi \over r p}\right)~.
\ee
which resembles the result for $S^3$ computed in \cite{Witten:1988hf}.

We now proceed to compute the all-loop gravitational partition function.  To do so we need to isolate those terms in \eqref{fullCSW} which come from the flat connections with holonomy
\be
(n_+,n_-)=\left({q+1\over 2},{q-1\over 2}\right)~,
\ee
for $A_\pm$. 

Interpreting the topological invariant as an infinite series of perturbative corrections is delicate and requires some discussion.  The path integral is an exact polynomial in powers of $r$, but not in powers $k_{cs}$.  In taking the semiclassical limit and reading off loop contributions in \eqref{fullCSW}, the shift by 2 in \eqref{rk} creates an potential ambiguity. Here we will use the one-loop result computed in the gravitational theory to give a precise dictionary between gravity and Chern-Simons theory.  In  the limit $r\to \infty$ the topological invariant \eqref{fullCSW} reproduces the  tree level and one-loop gravitational result \eqref{Z1lens} if  we identify the gravitational coupling as the analytic continuation of $r$, rather than $k_{cs}$. Explicitly, the correct prescription to reproduce the gravitational results is to take two copies of the Chern-Simons invariant with the identification
\be
r_+=-ik ~,\quad r_-=ik~.
\ee
This modified analytic continuation procedure is required to obtain the gravitational interpretation of the Chern-Simons path integral. Using this dictionary, the full gravitational path integral for $L(p,q)$ is
\bea\label{lensall}
Z_{\rm lens}&=&
{1\over  4kp}e^{2\pi k/p}\Bigg[ -2\cos({2\pi\over p})\left(\cos({2\pi \over p}q)+\cos({2\pi \over p}q^*)\right)+\cr&&+ e^{-{2\pi\over kp}}\left(\cos({4\pi\over p})+\cos({2\pi\over p}(q-q^*))\right) +e^{2\pi\over kp}\left(1+\cos({2\pi\over p}(q+q^*))\right)\Bigg]~.
\eea

Now that we have an expression at all orders in perturbation theory, we can consider the sum over geometries and attempt to regulate the sum.  The discussion is nearly identical to that in section 4.2. Using \eqref{lensall} in \eqref{Zlensfinal} we encounter again a divergent sum that cannot be regulated. In particular, the $p$ dependence of the higher loop terms does not fall-off quickly enough to make the sum convergent.  Moreover, the divergence can not be regulated using zeta function techniques.  Explicitly, if we perform the sum over $q$ and $p$ in \eqref{lensall} we get
\bea
Z_{\rm lens}&=&
\sum_{p=1}^\infty {1\over 4 kp}e^{2\pi k/p}\Bigg[-4\cos({2\pi\over p})\mu(p)+\cr && 
~~~~~~~ e^{-{2\pi\over kp}}\big(\cos({4\pi\over p})\phi(p)+S(1,-1,p)\big) +e^{2\pi\over kp}\big(\phi(p)+S(1,1,p)\big)\Bigg]
\eea
The divergent contributions to the sum are those proportional to $\phi(p)$.  At large $p$  the higher loop corrections proportional to $e^{2\pi \over kp}$ are irrelevant  and the computation is identical to that discussed below \eqref{reg}.  The result is a divergence proportional to $\zeta(1)$ which remains after zeta function regularization.

\section{Discussion and Speculation}

We have initiated a systematic study of quantum gravity in three dimensional de Sitter space, constructing explicitly a path integral including quantum gravitational effects due to both loops and instantons. 
We now discuss possible implications of our results.

\subsection{The Status of de Sitter Quantum Gravity}

We have computed the sum over a class of smooth Euclidean saddles of dS$_3$ gravity.  From our analysis we find that the sum is divergent and cannot be regulated using normal techniques. Explicitly, using zeta function regularization the path integral is
\be
Z = 24 \zeta(1) +\dots
\ee 
where $\dots$ denotes finite terms.  
This divergence is due to the infinite number of saddles with small volume in Planck units.  

This is in contrast to the AdS case.  In that case the corresponding divergence exists, but it can be regulated.  The corresponding AdS geometries have an important physical interpretation as those responsible for the modular invariance of the dual CFT.  Although the locally de Sitter geometries we have identified have a similar physical interpretation, we see no way to regulate the sum in this case.  We now discuss possible physical implications of this fact.

One possible conclusion is that quantum gravity in de Sitter space does not exist.  All known de Sitter vacua in string theory are unstable, due to either classical or quantum mechanical instabilities.  This may indicate a fundamental obstruction to de Sitter quantum gravity.  However, as long as the decay rate is slow compared to the Hubble time some patch of the space-time will inflate eternally, so it seems that eternally inflating backgrounds are generic in string theory. It would be surprising if quantum gravity theories could be defined in complicated eternally inflating backgrounds with bubble nucleation, but not in the highly symmetric de Sitter background.  

A second, related possibility is that  pure Einstein gravity is pathological in some way, but that other more complicated theories of de Sitter quantum gravity do exist.  As we saw in section 4, loop corrections due to gravity will suppress higher order terms in the sum over geometries.  In the case of pure gravity, this suppression was not sufficient to make the path integral converge.   However, it is easy to imagine that theories with more interesting matter content -- such as that coming from string theory -- will lead to further suppressions which render the sum convergent.  Unfortunately, a complete computation of the path integral is much more difficult once local degrees of freedom are included.  Our classification of classical solutions relied on the fact that all solutions to the equations of motion are locally de Sitter.  This is no longer true once matter fields with local degrees of freedom are present. One simple case where computations do seem possible  is three dimensional topologically massive gravity.  It may be that topologically massive gravity is the only purely metric theory of de Sitter gravity which can be defined consistently in three dimensions \cite{CLM2011}.  For a discussion of somewhat similar results in the AdS context see \cite{Maloney:2009ck}. 

Another possibility is that the divergence must be regulated in some manner.   One indication that this might be the case is that the divergence is in fact independent of the coupling constant $k$.  
This means that if we compute for example the expectation value of the Euclidean volume 
\be
\langle V \rangle = {{\partial_k Z}\over Z}. 
\ee
the numerator does not depend on the cutoff as the divergence in $Z$ is independent of $k$.  However, the denominator in this expression is, strictly speaking, infinite.  This can be remedied, for example, if we choose to sum only up to quotients with $|\Gamma| < \Lambda$ where $\Lambda$ is some cutoff of order $k$.  This removes saddles with sub-Planckian volume.  In order to make this proposal precise, however, one would need to demonstrate that the appropriate low energy observables are regulator independent.  

In the computations of this paper we have included only the sum over lens spaces.  It is possible that the divergence could disappear if additional saddle points (corresponding to $S^3/\Gamma$ where $\Gamma$ is non-abelian) are included in the sum.  
These other saddles do not have a simple Lorentzian interpretation so it is not clear a priori whether they should be included.  However, the question of which saddles should be included can be answered only if we give a precise definition to the quantum mechanical path integral, which we have not attempted.  If included these saddles will lead to additional divergences which might render the whole path integral finite.  This computation is easy to perform at tree level, as outlined in appendix A.  In this case we find that the additional geometries do not help.  One could also perform perturbative computations for these other saddles; this is a straightforward but difficult task.  We hope to turn to this in future work.  An intriguing possibility is that the sum will be finite only for certain (discrete) values of $k$, indicating that the quantum theory exists only for certain quantized values of the cosmological constant.  In the AdS case this is essentially what happens, since  the corresponding sum over geometries can only be regulated for certain values of $c_L$ and $c_R$.

A final possibility is that quantum gravity in de Sitter space makes sense, but we are not computing the correct physical quantity. For example, it might be that the correct definition of the path integral involves fixing boundary data of some sort.  In AdS/CFT the canonical ensemble partition function is given by a Euclidean path integral over geometries which are asymptotically a torus with fixed conformal structure.  This computation naturally leads to an interpretation of the modular group as bulk diffeomorphisms which act nontrivially at infinity.  In the case of de Sitter space there is no obvious definition of boundary, hence it is not clear what should be held fixed and how to identify the conformal structure relevant for the modular sum over $L(p,q)$. We hope future work will shed some light in this direction.     
 
 \subsection{Speculations on dS/CFT, Entropy and the Wave Function of the Universe}
 
We conclude with a few speculations on the relationship between our de Sitter Farey tail and other approaches to de Sitter quantum gravity, as well as the question of the nature and interpretation of the entropy of the cosmological horizon \cite{Gibbons:1977mu}.

We have argued that the path integral of de Sitter gravity naturally includes a sum over a coset of $SL(2,\Z)$.  This group is familiar as the group of modular transformations acting on the conformal structure parameter $\tau$ of a torus.  It is therefore natural to ask whether this modular group has an interpretation in terms of a two dimensional CFT.  Indeed, the dS/CFT conjecture states that three dimensional de Sitter gravity is related to a two dimensional Euclidean conformal field theory \cite{Strominger:2001pn}.  However, the precise relationship between our modular sum and the proposed CFT${}_2$ dual to dS${}_3$ is far from clear.
The dS/CFT conjecture is motivated by the observation that the group of bulk diffeomorphisms acts naturally as conformal transformations on the asymptotic boundary $\scri^\pm$ of de Sitter space.  Our sum arises in Euclidean signature where there is no boundary.  If anything, our sum appears to be related to a sum over conformal structures on the horizon of de Sitter space, rather than on $\scri^\pm$.  

We emphasize that the question of modular invariance of dS/CFT is fundamentally related to the problem of the de Sitter entropy. Perhaps the most notable success of the dS/CFT correspondence is the derivation of de Sitter entropy given by \cite{Park:1998qk, Bousso:2001mw,Balasubramanian:2002zh}.  This derivation starts by assuming that the CFT dual to de Sitter space is modular invariant, so that Cardy's formula can be used to compute the asymptotic density of states.  The answer matches precisely the Bekenstein-Hawking  entropy of the de Sitter horizon.  Cardy's formula applies only to unitary CFTs with a normalizable ground state.  Thus without any further understanding of the CFT dual to de Sitter space this result should be regarded as suggestive, but not as a complete derivation of de Sitter entropy.
 
 It is tempting to speculate that our lens spaces are the bulk geometries responsible for the modular invariance of the CFT${}_2$ dual to de Sitter space.   This is precisely how it works in the AdS case; the sum over bulk geometries related by $SL(2,\Z)$ modular transformations naturally leads to the modular invariance of the CFT partition function.  A more careful study of the rotation from Lorentzian to Euclidean signature might shed light on this issue.\footnote{As an interesting aside, we note that if the sum over geometries is to include the sum over all elements of $\Z \backslash SL(2,\Z)/ \Z$, we should include the lens spaces $L(p,q)$ for all coprime values of $(p,q)$.  This includes the case $(p,q)=(0,1)$, for which $L(0,1)=S^1\times S^2$.  This is not a smooth saddle point of the Euclidean equations of motion; instead it is a singular solution with zero action.  Our conjecture is that the proper Euclidean path integral should include this singular saddle as well.}
 
We conclude with the observation that there is a slightly different setting in which the modular invariance of dS/CFT might arise.  Rather than the Euclidean path integrals considered in this paper, one could imagine computing the wave function of the universe via a Lorentzian path integral of the form 
\be\label{psidef}
\psi(h) \sim \int_{g|_{\partial M} = h} {\cal D} g~e^{i k S}~.
\ee  
This path integral is a functional of the induced metric $h$ on a two dimensional space-like slice, as well as on some (unspecified) initial data which determine the state.  The dS/CFT correspondence is the statement that as the spacelike slice is taken towards $\scri^+$, $\psi(h)$ will approach the partition function of a two dimensional CFT, regarded as a function of the conformal structure of the spacelike slice.  When the space-like slice is taken to be a sphere the dominant contribution comes from the usual de Sitter geometry and the wave function can be computed explicitly.  This computation was described in \cite{Maldacena:2002vr}.
  
However, one could also take spacelike slice in the wave function (\ref{psidef}) to be a torus at timelike future infinity.  In this case $\psi$ is conjectured to be a CFT partition function on a torus, which should exhibit the expected modular invariance.  Although we will not attempt to compute the wave function explicitly, it is easy to see how this modular invariance should arise.  The saddle point geometry which will arise in the semiclassical approximation to (\ref{psidef}) is not de Sitter space, but rather the quotient dS${}_3/\Z$.  This can be understood by writing the metric on  de Sitter space in cyclindrical coordinates 
\be\label{tmetric}
ds^2 = -dt^2 + \cosh^2t d\theta^2 + \sinh^2 t d\phi^2~,
\ee
with $\phi\sim\phi+2\pi$.  We can further identify $\theta\sim\theta+2\pi$ to obtain a geometry which approaches a torus at future infinity $t\to\infty$. This quotient $dS_3/ \Z$ has a singularity at $t=0$.  This geometry is a saddle point which will contribute to the wave function (\ref{psidef}).  

In fact there are an infinite number of such saddle point geometries, related by modular transformations, which are labelled by the coset $SL(2,\Z)/\Z$.  To see this, note that the $\phi$ and $\theta$ coordinates are not treated democratically in the geometry (\ref{tmetric}).  Indeed, one of them shrinks to zero size at the Milne singularity $t=0$ whereas the other has finite size.  Thus in writing the saddle (\ref{tmetric}) we have singled out one of the two cycles of the boundary torus.  By a change of coordinates, one can obtain a geometry where any cycle of the boundary torus shrinks to zero, not just the $\theta$ cycle.  These geometries are related by large diffeomorphisms in the bulk, and hence by large conformal transformations on the boundary.   These large conformal transformations are modular $SL(2,\Z)$ transformations and the corresponding geometries are labelled the coset $SL(2,\Z)/\Z$.  We expect that this set of modular transformations is related by analytic continuation to the modular sum over lens spaces described in this paper; it would be nice to make this correspondence explicit.

\section*{Acknowledgements}

We are grateful to Raphael Bousso, Steve Carlip, Mike Douglas, Sergei Gukov, Sean Hartnoll, Chris Herzog, Diego Hofman, Simone Giombi, Leonard Susskind, James Sully, Andy Strominger, Edward Witten and Xi Yin for useful conversations.  This work is supported by the National Science and Engineering Research Council of Canada.

\appendix

\section{Classical Saddle Point Contributions}\label{quo}

In this appendix we compute the tree-level contributions to the path integral of all smooth solutions to the Euclidean equations of motion. As discussed in section 3 these classical saddles are quotients $S^3/\Gamma$ where $\Gamma$ is a discrete, freely acting subgroup of $SO(4)$.  Their classical contribution to the action is given by \eqref{c:tree}. 
With the exception of the lens spaces described in section 2, the groups $\Gamma$ are non-abelian and are central extensions of crystallographic groups.  Such spaces are uniquely labelled by their fundamental group $\Gamma$.  We refer the reader to \cite{thurston97} for a complete classification.

\subsection*{Dihedral case} 

When $\Gamma$ is a central extension of the dihedral group the spherical manifold is known as a prism manifold. The fundamental group is
\be
\langle x, y\, |\, x^{-1}yx=y^{-1}, x^{2m}=y^n\rangle~, 
\ee
with $m\ge 1$ and $n\ge 2$ and $\Gamma$ is of order $4mn$. The tree level sum over geometries is 
\bea
Z_{\rm prism}^{(0)} &=& \sum_{m=1\atop n=2}^{\infty}e^{2\pi k/(4mn)} ~.
\eea
Taylor expanding the exponential gives 
\bea\label{ccb}
Z_{\rm prism}^{(0)}&=& \sum_{r=0}^{\infty}\left({\pi k\over 2}\right)^{r}{1\over r!}\left(\sum_{m=1}^{\infty} m^{-r} \right)\left(\sum_{n=2}^{\infty} n^{-r}\right)\nonumber\\
&=&\sum_{r=0}^{\infty} \left({\pi k\over 2}\right)^{r}{1\over r!} \,{\zeta(r)(\zeta(r)-1)}~.
\eea
In this case $Z_{\rm prism}^{(0)}$ has a simple and double pole at $r=1$. 

\subsection*{Tetrahedral case} 
In this case the fundamental group can take one of two forms.  It is a product of cyclic group of order $m$ with either a binary tetrahedral group of order $24$ or a general tetrahedral group of $8\cdot 3^n$ with $n\ge 1$. In both cases $m$ is coprime to 6.
For the binary tetrahedral case, $\Gamma$ is of order $24 m$ and the tree level partition sum is
\bea\label{cx}
Z_{\rm bi-tetra}^{(0)} &=& \sum_{m=1\atop (m,6)=1}^\infty e^{2\pi k/(24 m )} \nonumber\\
&=& \sum_{r=0}^{\infty}\left({\pi k\over 12}\right)^{r}{1\over r!}\sum_{m=1\atop(m,6)=1}^{\infty} m^{-r} \\
&=&\sum_{r=0}^{\infty} \left({\pi k\over 12}\right)^{r}{1\over r!} \,\zeta(r){(1-2^{-r})(1-3^{-r})}~.\nonumber
\eea
In the general case $|\Gamma|=8\cdot 3^{n}m$ and the contribution to $Z^{(0)}$ is 
\bea\label{cxa}
Z_{\rm tetra}^{(0)} &=& \sum_{m=1\atop (m,6)=1}^\infty\sum_{n= 1}^{\infty}e^{2\pi k/(8 m 3^{n})} \nonumber\\
&=& \sum_{r=0}^{\infty}\left({\pi k\over 4}\right)^{r}{1\over r!}\sum_{m=1\atop(m,6)=1}^{\infty} m^{-r} \sum_{n=1}^{\infty} 3^{-r n}\\
&=&
\sum_{r=0}^{\infty}  \left({\pi k\over 12}\right)^{r}{1\over r!}\,\zeta(r){(1-2^{-r})}~.\nonumber
\eea
Here both sums have a pole at $r=1$. 

\subsection*{Octahedral case}

Here $\Gamma$ is a product of the cyclic group of order $m$ with the binary octahedral group of order $48$, so $|\Gamma|=48m$.  The order $m$ must be coprime to 6  so the contribution to the partition function is
\bea\label{cf}
Z_{\rm oct}^{(0)} &=& \sum_{m=1\atop (m,6)=1}^\infty e^{2\pi k/(48 m )} \nonumber\\
&=& \sum_{r=0}^{\infty}\left({\pi k\over 24}\right)^{r}{1\over r!}\sum_{m=1\atop(m,6)=1}^{\infty} m^{-r} \\
&=&\sum_{r=0}^{\infty} \left({\pi k\over 24}\right)^{r}{1\over r!}\, \zeta(r){(1-2^{-r})(1-3^{-r})}~.\nonumber
\eea
 
\subsection*{Icosahedral case} 

For the last class of spherical manifolds the fundamental group is a product of a cyclic group of order $m$ coprime to 30 with the binary icosahedral group.  Here $\Gamma$ is of order $120m$ and the saddle contribution is
\bea\label{cfa}
Z_{\rm icos} ^{(0)}&=& \sum_{m=1\atop (m,30)=1}^\infty\sum_{n= 1}^{\infty}e^{2\pi k/(120 m )}\nonumber \\
&=& \sum_{r=0}^{\infty}\left({\pi k\over 60}\right)^{r}{1\over r!}\sum_{m=1\atop(m,30)=1}^{\infty} m^{-r} \\
&=&\sum_{r=0}^{\infty}\left({\pi k\over 60}\right)^{r}{1\over r!}\, \zeta(r){(1-2^{-r})(1-3^{-r})(1-5^{-r})}~.\nonumber
\eea

\subsection*{Putting it together}\label{quo:final}
Gathering the results from all five classes of $S^3/\Gamma$ geometries and adding  the classical saddle point contributions, the partition function at tree level is
\bea\label{tree}
Z^{(0)}&=&Z_{\rm lens}^{(0)}+Z_{\rm prism}^{(0)}+Z_{\rm bi-tetra}^{(0)}+Z_{\rm tetra}^{(0)}+Z_{\rm oct}^{(0)}+Z_{\rm icos} ^{(0)}~,
\eea
with each individual term given by \eqref{cca}, \eqref{ccb}, \eqref{cx}, \eqref{cxa}, \eqref{cf} and \eqref{cfa}. As it stands the sum \eqref{tree} is divergent, and the divergences are determined by the pole of the zeta function. In particular,  it has a double pole contained in $Z_{\rm prism}^{(0)}$ due to the term $\zeta(1)^2$ with coefficient 
\be
{\pi k\over 2}~.
\ee
In addition \eqref{tree} has a single pole due to $\zeta(1)$ in all six saddles in \eqref{tree} and the residue is
\bea
12k^2-{371\over 30^2}\pi k~.
\eea
We conclude that the classical partition function diverges even after zeta function regularization. 


\section{Details on One-loop Determinants}

\subsection{Harmonic decomposition}\label{app:Harmonic}

In section \ref{heat}, and in particular to obtain \eqref{detloop}, we use the harmonic decomposition for vectors and 2-tensors. For completeness we review this decomposition here.  

A vector can be split into transverse and longitudinal modes
\be
V_\mu=T_\mu+L_\mu~,
\ee
where 
\be\label{LT}
\nabla^\mu T_\mu=0~, \quad L_\mu=\nabla_\mu \phi~,
\ee   
with $\phi$ a scalar. For a symmetric 2-tensor the analogous decomposition is
\be
h_{\mu\nu}=T_{\mu\nu}+{1\over D}g_{\mu\nu}\psi+L_{\mu\nu}~.
\ee
Here $T_{\mu\nu}$ is symmetric, transverse and traceless
\be
T_{\mu}^\mu=0~,\quad \nabla^\mu T_{\mu\nu}=0~,
\ee
and $L_{\mu\nu}$ is the longitudinal and traceless which we write as
\be\label{L}
L_{\mu\nu}=\nabla_{(\mu} V_{\nu)}-{2\over D}g_{\mu\nu}\nabla^\alpha V_\alpha~.
\ee
Note that the decomposition for $L_{\mu\nu}$ is not unique. The vector $V_\mu+C_\mu$ with $C_\mu$ a conformal Killing vector gives the same tensor $L_{\mu\nu}$. We can further split the vector in \eqref{L} into its harmonic components, breaking down $L_{\mu\nu}$ into longitudinal-transverse (LT)  and longitudinal-longitudinal (LL) components:
\be
L_{\mu\nu}^{LT}=\nabla_{(\mu} T_{\nu)}~,\quad L_{\mu\nu}^{LL}=\nabla_{(\mu} L_{\nu)}-{2\over D}g_{\mu\nu}\nabla^\alpha L_\alpha~,
\ee
with $T_\mu$ and $L_\mu$ as defined in \eqref{LT}.

\subsection{Heat kernel regularization}\label{app:HK}

Here we give a detailed derivation of the zeta function for lens spaces \eqref{ffzhk}. 
Starting from  \eqref{K1} and using \eqref{dnj}, we have
\bea\label{Z1hk}
\log Z^{(1)}&=&{1\over 2p}\sum_{m\in Z_p}\Bigg(\sum_{n=3}^\infty {\ln(n^2-4)\over  \sin{m\tau\over 2}\sin{m\bar\tau\over 2}}\left[\cos(m\tau_1)\cos(nm\tau_2)-\cos(m\tau_2)\cos(nm\tau_1)\right]\cr&&-\sum_{n=3}^\infty {\ln(n^2-1)\over  \sin{m\tau\over 2}\sin{m\bar\tau\over 2}}\left[\cos(2m\tau_1)\cos(nm\tau_2)-\cos(2m\tau_2)\cos(nm\tau_1)\right]\Bigg)~.
\eea

The sums we have to regulate are of the form
\be
\sum_{n=3} \ln(n^2-1)\cos(nx)~,\quad \sum_{n=3} \ln(n^2-4)\cos(nx)~,
\ee
with $x=m\tau_{1,2}$. A useful way to regulate such expressions is by defining\footnote{For a pair of functional operators $A$ and $B$ it is not true that regulated determinant $\det(AB)$ is equal to the product of the regulated determinants $\det(A) \times \det(B)$.  This ``anomaly" arises because the zeta functions associated with these operators might have poles with non-zero residue \cite{Quine}. We have checked explicitly that this anomaly does not arise for the $L(p,q)$ one loop determinants. Thus we can safely use the product formula $$\prod_k (\lambda^2_k-\lambda^2)=\prod_k (\lambda_k-\lambda)\prod_k (\lambda_k+\lambda)~.$$   }
\be
\zeta_1(s,x)=\sum_{n=3}^\infty{\cos(nx)\over (n-1)^s}+{\cos(nx)\over (n+1)^s}~,
\ee
and
\be
\zeta_2(s,x)=\sum_{n=3}^\infty{\cos(nx)\over (n-2)^s}+{\cos(nx)\over (n+2)^s}~.
\ee
The above functions satisfy
\bea
{d\over ds}\zeta_1(0,x)&=&-\sum_{n}\ln(n^2-1)\cos(nx)~,\cr{d\over ds}\zeta_2(0,x)&=&-\sum_{n}\ln(n^2-4)\cos(nx)~.
\eea
By shifting the sums in $\zeta_{1,2}(s,x)$ we get
\bea\label{z12}
\zeta_1(s,x)&=&2\cos(x)C(s,x)-\sum_{n=2}^{n=3}\cos((n-1)x)n^{-s}~,\cr
\zeta_2(s,x)&=&2\cos(2x)C(s,x)-\sum_{n=2}^{n=4}\cos((n-2)x)n^{-s}~,
\eea
where we dropped terms independent of $s$ and we defined 
\be
C(s,x)=\sum_{n=1}^\infty\cos(nx)n^{-s}~.
\ee

Using \eqref{z12} we construct the zeta function for the non-zero eigenvalues in \eqref{Z1hk},
\bea\label{zzhk}
\zeta(s)_{\rm lens}&=&{1\over 2p}\sum_{m\in Z_p} {1\over  \sin{m\tau\over 2}\sin{m\bar\tau\over 2}}\Big[\cos(m\tau_1)\zeta_2(s,m\tau_2)-\cos(m\tau_2)\zeta_2(s,m\tau_1)\cr 
&&\quad\quad\quad-\cos(2m\tau_1)\zeta_1(s,m\tau_2)+\cos(2m\tau_2)\zeta_1(s,m\tau_1)\Big]
\eea
The explicit relation between  $\zeta(s)_{HK}$ and $\log (Z^{(1)})$ is given in \eqref{Z1zhk}. Using \eqref{z12} we can simplify \eqref{zzhk} as
\bea\label{zzk}
\zeta(s)_{\rm lens}=&&{2\over p}\sum_{m\in \Z_p}(1+\cos(m\tau)+\cos(m\bar\tau))[C(s,m\tau_1)+C(s,m\tau_2)]-{1\over 4^s}\nonumber\\
&&-\frac{1}{p}\sum_{m=0}^{p-1}(\cos(m\tau)+\cos(m\bar\tau))\left(\frac{1}{2^s}+\frac{1}{4^s}\right),
\eea
The advantage of working with \eqref{zzk} is that now the sum over $m$ is straightforward. For example, consider 
\bea\label{cos}
&&\sum_{n=1}^\infty\sum_{m=0}^{p-1}\frac{1}{n^s}\cos(m\tau)\cos(mn\tau_1)=
\cr&&= {1\over 2}\sum_{n=1}^\infty\sum_{m=0}^{p-1}\frac{1}{n^s}\left[\cos\left(2\pi {mq\over p}(q^*-1+n)\right)+\cos\left(2\pi {mq\over p}(q^*-1-n)\right)\right]\cr
&&={p^{1-s}\over 2}\left[\zeta(s,-{q^*-1\over p})+\zeta(s,{q^*-1\over p})-\left({p\over 1-q^*}\right)^s\right]~,
\eea
where in the second line of \eqref{cos} we introduced the $q^*$ which satisifies $qq^*=1\mod p$, and in the last line we used 
\be
\zeta(s,a)=\sum_{n=0}^\infty{1\over (n+a)^s}~,
\ee
the zeta Hurwitz function. It is important to note that the derivation \eqref{cos} is not valid if $q=\pm 1\mod p$ and it will be a case we will treat separately below. Assuming $q\neq\pm 1\mod p$ and implementing \eqref{cos} in all terms in \eqref{zzk} we get
\bea\label{fzhk}
&&\zeta(s)_{\rm lens}=p^{-s}\sum_{\pm}\Bigg[\zeta(s,\pm{q^*-1\over p})+\zeta(s,\pm{q^*+1\over p})+\zeta(s,\pm{q-1\over p})+\zeta(s,\pm{q+1\over p})\cr 
&&\quad \quad\quad-\left({p\over \pm 1-q^*}\right)^s-\left({p\over \pm1- q}\right)^s\Bigg]+4p^{-s}\zeta(s)-{1\over 4^s}~.
\eea

We now consider the case $q=\pm 1\mod p$. From \eqref{zzk} we have
\bea\label{zhkq1}
\zeta(s)_{(p,1)}&=&\zeta(s)_{(p,p-1)}\cr &=&{4\over p}\sum_{m\in \Z_p}\left[2+\cos\left({4\pi m\over p}\right)\right]C\left(s, {2\pi m\over p}\right)-{2\over 4^s}-{1\over 2^s}\cr
&=&2p^{-s}\left[\zeta(s,{2\over p})+\zeta(s,-{2\over p})-\left(-{p\over 2}\right)^s\right]+8p^{-s}\zeta(s)-{2\over 4^s}-{1\over 2^s}~,
\eea
which is valid for $p>2$. The special case $p=2$ and $q=1$ gives
\bea\label{zhkq2}
\zeta(s)_{(2,1)}&=&{6}\sum_{m\in \Z_2}C\left(s, {\pi m}\right)-{3\over 4^s}-{2\over 2^s}\cr
&=&{12\over 2^{s}}\zeta(s)-{3\over 4^s}-{2\over 2^s}~.
\eea

\section{Dirichlet Series and Related Formulas}\label{app:formulas}

In this appendix we summarize some useful number-theoretic formulae.

\subsection{Riemann and Related Zeta Functions}

\noindent{\it Riemann zeta function:} 

The Riemann zeta function $\zeta(s)$ is the analytic continuation of the series 
\be
\zeta(s)=\sum_{n=1}^\infty{1\over n^s}=\prod_{p \,{\rm prime}}(1-p^{s})^{-1} ~,
\ee
to the complex $s$ plane.  
The function has a simple pole at $s=1$ and Laurent series
\be
\zeta(s)={1\over s-1}+\sum_{k=0}^\infty\gamma_k{(-1)^k\over k!}(s-1)^k
\ee
where $\gamma_k$ is the Stieltjes constant. Some useful values of $\zeta(s)$ are
\be\label{zeta}
\zeta(0)=-{1\over 2}~,\quad {d\over ds}\zeta(0)=-{1\over 2}\ln (2\pi)
\ee

\noindent {\it Hurwtiz zeta function:}

A simple generalization of the Riemann zeta function is the Hurwitz function 
\be
\zeta(s,a)=\sum_{n=0}^\infty{1\over (n+a)^s}~.
\ee
It is a meromorphic function in $s$ and ${\Re}(a)>-1$ with a simple pole at $s=1$.  We will need the following values
%
\be\label{hurwitz}
\zeta(0,a)={1\over 2}-a~,\quad {d\over ds}\zeta(0,a)=\ln(\Gamma(a))-{1\over 2}\ln (2\pi)~.
\ee
so that in particular
\be\label{gg}
{d\over ds}\zeta(0,a)+{d\over ds}\zeta(0,-a)=-\ln(\sin(\pi a))-\ln(-2a)~.
\ee
%

\noindent {\it Euler's totient function:}

The Euler's totient function $\phi(p)$ is defined as the number of positive integers less than $p$ which are relatively prime to $p$.  The Dirichlet series for the totient function is
\be\label{app:euler}
\sum_{n=1}^\infty\phi(n)n^{-s}={\zeta(s-1)\over \zeta(s)}~.
\ee

\noindent {\it  Ramanujan's sum and Mobius function:}

For a pair of integers $m$ and $s$, Ramanujan's sum is defined as
\be\label{Ram}
c_m(s)=\sum_{n=1\atop (m,n)=1}^m \exp\left(2\pi i {n\over m} s\right)~.
\ee
The Mobius function is defined as
\be
\mu(m)=\left\{ \begin{array}{ll} 0 & \textrm{if  $m$ has one or more repeated prime factor,}\\
 1 & \textrm{ if $m=1$,}\\ (-1)^k & \textrm{ if $m$ is a product of $k$ distinct primes} \end{array} \right.
\ee
It  satisfies 
\be
\mu(m)=c_m(1)~,
\ee
The Dirichlet series for the Mobius function is
\be\label{app:mu}
\sum_{m=1}^\infty m^{-s}\mu(m)={1\over \zeta(s)}
\ee
This has no poles for positive integer values of $s$.

\noindent {\it Dedekind sum:}

For a pair of coprime integers $(c,d)$ with $c>0$, the Dedekind sum is defined by
\begin{equation}
s(d,c)=\frac{1}{4c}\sum_{k=1}^{c-1}\cot\frac{\pi k}{c}\cot\frac{\pi d k}{c}~.
\end{equation}

\subsection{Kloosterman Zeta Function}\label{A:Kloo}
We now summarize a few features of  Kloosterman zeta functions. The Kloosterman sum is defined as
\be
S(a,b,m)=\sum_{n=1\atop (n,m)=1}^m\exp(2\pi i(an^*+bn)/m)
\ee
where $n^*$ is the inverse of $n$ modulo $m$. 
We are interested in sums of the form 
\be
L(m,n;s) = \sum_{p=1}^\infty p^{-2s} S(m,n;p)~.
\ee
This is known as the Kloosterman zeta function.   This series converges absolutely when $\Re s >1/2$.  The structure of $L(m,n,s)$ is quite rich, and its poles contain data about the spectrum of the hyperbolic Laplacian on $\H/SL(2,\Z)$. We will summarize a few of its salient features here, focusing primarily on its analytic properties on the real $s$ axis, and refer the reader to \cite{Iwaniec:2002aa} for details and proofs.

We first consider the simple case where either $m$ or $n$ is equal to zero.  In this case the Kloosterman sum reduces to a Ramanujan sum and
\be
L(0,n;s)={1\over n^{2s}\zeta(2s)}\sum_{\delta | n}\delta^{1-2s}~,
\ee
has no poles and is analytic everywhere.  Moreover, $L(0,n;s)$ vanishes at $s=1/2$.

The analytic properties are most conveniently summarized by the function  \cite{Iwaniec:2002aa} 
\be
Z(m,n;s) = {1\over 2 \sqrt{mn}}\sum_{p=1}^\infty p^{-1} S(m,n;p) J_{2s-1} ({4  \pi \over p} \sqrt{mn})~,
\ee
when $mn$ positive, with a similar formula for $mn$ negative.  Using the Neumann expansion
\be
z^{\nu}=2^\nu \sum_{k=0}^\infty {(\nu+2k)\Gamma(\nu+k)\over k!} J_{\nu+2k}(z)
\ee
with $z={4  \pi \over p} \sqrt{mn}$ and $\nu=2s-1$
we see that
\be
L(m,n;s) = {2^{2s} \sqrt{mn}\over (4 \pi \sqrt{mn})^{2s-1}}\sum_{k=0}^\infty {(2(s+k)-1)\Gamma(2s-1+k)\over k!}   Z(m,n;s+k)~.
\ee

The functions $L(m,n;s)$ and $Z(m,n;s)$ can both be analytically continued to meromorphic functions on the complex $s$ plane.  The locations of the poles are related to the eigenvalues of the hyperbolic Laplacian on $\H/SL(2,\Z)$.  This operator has a discrete spectrum which is bounded below at $1/4$.  We denote one of the eigenvalues as $\lambda_j=-s_j (s_j-1)$, where $s_j=1/2+i t_j$.  We note that when $\lambda >1/4$ the $t_j$ are real.  These eigenvalues lead to simple poles for $Z(m,n;s)$ at $s=s_j$.  These poles do not concern us, as they are away from the $\Re s$ axis.  The remaining possible pole is at $s=1/2$.  If the hyperbolic Laplacian has an eigenvalue $1/4$, then this would lead to a double pole at $s=1/2$.  However, for $SL(2,\Z)$ the first eigenvalue appears at $\lambda>1/4$ so there is no double pole.  There is, however, the possibility of a simple pole at $s=1/2$, even without an eigenvalue at $\lambda=1/4$:
\be
Z(m,n;s)\sim {R(m,n)\over s-1/2} + \dots~,
\ee
However, the residue of the pole at $s=1/2$ was computed in  \cite{Iwaniec:2002aa}  to be
\be
R(m,n) = {-1\over 4} \phi(m,1/2)\phi(n,1/2)~,
\ee
where 
\be
\phi(n,s)= {\pi^s \over \Gamma(s)} |n|^{s-1} L(0,n;s)~,
\ee
is zero at $s=1/2$.  We conclude that $Z(m,n;s)$ has no poles on the real $s$ axis.  

From this it follows that the only poles of $L(m,n;s)$ on the real $s$ axis come from the gamma function, which has simple poles at the non-positive integers.  For the $k=0$ term in the sum these poles are cancelled by the coefficient $2(s+k)-1$.  Thus $L(m,n;s)$ has no pole at $s=1/2$.  However, when $s=-n/2$, $n=0,1,\dots$ there will be simple poles.  For example, there is a pole at $s=0$ with non-zero residue coming from the $k=1$ term:
\be
L(m,n;s)\sim {1\over s} 4 \pi mn Z(m,n;1) + \dots~.
\ee
Similar conclusions hold for the case where $mn$ is negative.

\bibliographystyle{utphys}
\bibliography{dsref}

\providecommand{\href}[2]{#2}\begingroup\raggedright\begin{thebibliography}{10}

\bibitem{Witten:2001kn}
E.~Witten, ``{Quantum gravity in de Sitter space},''
\href{http://arXiv.org/abs/hep-th/0106109}{{\tt hep-th/0106109}}.

\bibitem{Strominger:2001pn}
A.~Strominger, ``{The dS/CFT correspondence},'' {\em JHEP} {\bf 10} (2001) 034,
\href{http://arXiv.org/abs/hep-th/0106113}{{\tt hep-th/0106113}}.

\bibitem{Maldacena:2002vr}
J.~M. Maldacena, ``{Non-Gaussian features of primordial fluctuations in single
  field inflationary models},'' {\em JHEP} {\bf 05} (2003) 013,
\href{http://arXiv.org/abs/astro-ph/0210603}{{\tt astro-ph/0210603}}.

\bibitem{Goheer:2002vf}
N.~Goheer, M.~Kleban, and L.~Susskind, ``{The trouble with de Sitter space},''
  {\em JHEP} {\bf 07} (2003) 056,
\href{http://arXiv.org/abs/hep-th/0212209}{{\tt hep-th/0212209}}.

\bibitem{Kachru:2003aw}
S.~Kachru, R.~Kallosh, A.~D. Linde, and S.~P. Trivedi, ``{De Sitter vacua in
  string theory},'' {\em Phys. Rev.} {\bf D68} (2003) 046005,
\href{http://arXiv.org/abs/hep-th/0301240}{{\tt hep-th/0301240}}.

\bibitem{Hartle:1983ai}
J.~B. Hartle and S.~W. Hawking, ``{Wave Function of the Universe},'' {\em Phys.
  Rev.} {\bf D28} (1983)
2960--2975.

\bibitem{Dijkgraaf:2000fq}
R.~Dijkgraaf, J.~M. Maldacena, G.~W. Moore, and E.~P. Verlinde, ``{A black hole
  farey tail},''
\href{http://arXiv.org/abs/hep-th/0005003}{{\tt hep-th/0005003}}.

\bibitem{Maldacena:1998bw}
J.~M. Maldacena and A.~Strominger, ``{AdS(3) black holes and a stringy
  exclusion principle},'' {\em JHEP} {\bf 12} (1998) 005,
\href{http://arXiv.org/abs/hep-th/9804085}{{\tt hep-th/9804085}}.

\bibitem{Kraus:2006nb}
P.~Kraus and F.~Larsen, ``{Partition functions and elliptic genera from
  supergravity},'' {\em JHEP} {\bf 01} (2007) 002,
\href{http://arXiv.org/abs/hep-th/0607138}{{\tt hep-th/0607138}}.

\bibitem{Maloney:2007ud}
A.~Maloney and E.~Witten, ``{Quantum Gravity Partition Functions in Three
  Dimensions},'' {\em JHEP} {\bf 02} (2010) 029,
\href{http://arXiv.org/abs/0712.0155}{{\tt 0712.0155}}.

\bibitem{Witten:1988hc}
E.~Witten, ``{(2+1)-Dimensional Gravity as an Exactly Soluble System},'' {\em
  Nucl. Phys.} {\bf B311} (1988)
46.

\bibitem{Witten:1988hf}
E.~Witten, ``{Quantum field theory and the Jones polynomial},'' {\em Commun.
  Math. Phys.} {\bf 121} (1989)
351.

\bibitem{Achucarro:1987vz}
A.~Achucarro and P.~K. Townsend, ``{A Chern-Simons Action for Three-Dimensional
  anti-De Sitter Supergravity Theories},'' {\em Phys. Lett.} {\bf B180} (1986)
89.

\bibitem{Achucarro:1989gm}
A.~Achucarro and P.~K. Townsend, ``{Extended supergravities in d = (2+1) as
  Chern-Simons Theories},'' {\em Phys. Lett.} {\bf B229} (1989)
383.

\bibitem{Gukov:2003na}
S.~Gukov, ``{Three-dimensional quantum gravity, Chern-Simons theory, and the
  A-polynomial},'' {\em Commun. Math. Phys.} {\bf 255} (2005) 577--627,
\href{http://arXiv.org/abs/hep-th/0306165}{{\tt hep-th/0306165}}.

\bibitem{Jeffrey:1992tk}
L.~C. Jeffrey, ``{Chern-Simons-Witten invariants of lens spaces and torus
  bundles, and the semiclassical approximation},'' {\em Commun. Math. Phys.}
  {\bf 147} (1992)
563--604.

\bibitem{Carlip:1992wg}
S.~Carlip, ``{The Sum over topologies in three-dimensional Euclidean quantum
  gravity},'' {\em Class. Quant. Grav.} {\bf 10} (1993) 207--218,
\href{http://arXiv.org/abs/hep-th/9206103}{{\tt hep-th/9206103}}.

\bibitem{Guadagnini:1995wv}
E.~Guadagnini and P.~Tomassini, ``{Sum over the geometries of three
  manifolds},'' {\em Phys. Lett.} {\bf B336} (1994)
330--336.

\bibitem{Banados:1998tb}
M.~Banados, T.~Brotz, and M.~E. Ortiz, ``{Quantum three-dimensional de Sitter
  space},'' {\em Phys. Rev.} {\bf D59} (1999) 046002,
\href{http://arXiv.org/abs/hep-th/9807216}{{\tt hep-th/9807216}}.

\bibitem{Park:1998yw}
M.-I. Park, ``{Symmetry algebras in Chern-Simons theories with boundary:
  Canonical approach},'' {\em Nucl.Phys.} {\bf B544} (1999) 377--402,
  \href{http://arXiv.org/abs/hep-th/9811033}{{\tt hep-th/9811033}}.

\bibitem{Govindarajan:2002ry}
T.~R. Govindarajan, R.~K. Kaul, and V.~Suneeta, ``{Quantum gravity on dS(3)},''
  {\em Class. Quant. Grav.} {\bf 19} (2002) 4195--4205,
\href{http://arXiv.org/abs/hep-th/0203219}{{\tt hep-th/0203219}}.

\bibitem{Deser:1983nh}
S.~Deser and R.~Jackiw, ``{Three-Dimensional Cosmological Gravity: Dynamics of
  Constant Curvature},'' {\em Annals Phys.} {\bf 153} (1984) 405--416.

\bibitem{Parikh:2002py}
M.~K. Parikh, I.~Savonije, and E.~P. Verlinde, ``{Elliptic de Sitter space:
  dS/Z(2)},'' {\em Phys. Rev.} {\bf D67} (2003) 064005,
\href{http://arXiv.org/abs/hep-th/0209120}{{\tt hep-th/0209120}}.

\bibitem{Murthy:2009dq}
S.~Murthy and B.~Pioline, ``{A Farey tale for N=4 dyons},'' {\em JHEP} {\bf
  0909} (2009) 022, \href{http://arXiv.org/abs/0904.4253}{{\tt 0904.4253}}.

\bibitem{thurston97}
{Thurston, William P.}, {\em {Three-Dimensional Geometry and Topology}}.
\newblock {Princeton University Press}, Jan., {1997}.

\bibitem{Manschot:2007ha}
J.~Manschot and G.~W. Moore, ``{A Modern Fareytail},'' {\em Commun. Num. Theor.
  Phys.} {\bf 4} (2010) 103--159,
\href{http://arXiv.org/abs/0712.0573}{{\tt 0712.0573}}.

\bibitem{Witten:2010cx}
E.~Witten, ``{Analytic Continuation Of Chern-Simons Theory},''
\href{http://arXiv.org/abs/1001.2933}{{\tt 1001.2933}}.

\bibitem{Gibbons:1978ji}
G.~W. Gibbons and M.~J. Perry, ``{Quantizing Gravitational Instantons},'' {\em
  Nucl. Phys.} {\bf B146} (1978)
90.

\bibitem{Christensen:1979iy}
S.~M. Christensen and M.~J. Duff, ``{Quantizing Gravity with a Cosmological
  Constant},'' {\em Nucl. Phys.} {\bf B170} (1980)
480.

\bibitem{Yasuda:1983hk}
O.~Yasuda, ``{ON THE ONE LOOP EFFECTIVE POTENTIAL IN QUANTUM GRAVITY},'' {\em
  Phys. Lett.} {\bf B137} (1984)
52.

\bibitem{Gibbons:1978ac}
G.~W. Gibbons, S.~W. Hawking, and M.~J. Perry, ``{Path Integrals and the
  Indefiniteness of the Gravitational Action},'' {\em Nucl. Phys.} {\bf B138}
  (1978)
141.

\bibitem{Polchinski:1988ua}
J.~Polchinski, ``{The Phase of the Sum Over Spheres},'' {\em Phys. Lett.} {\bf
  B219} (1989)
251.

\bibitem{David:2009xg}
J.~R. David, M.~R. Gaberdiel, and R.~Gopakumar, ``{The Heat Kernel on AdS$_3$
  and its Applications},'' {\em JHEP} {\bf 04} (2010) 125,
\href{http://arXiv.org/abs/0911.5085}{{\tt 0911.5085}}.

\bibitem{Hawking:1976ja}
S.~W. Hawking, ``{Zeta Function Regularization of Path Integrals in Curved
  Space-Time},'' {\em Commun. Math. Phys.} {\bf 55} (1977)
133.

\bibitem{Witten:2007kt}
E.~Witten, ``{Three-Dimensional Gravity Revisited},''
\href{http://arXiv.org/abs/arXiv:0706.3359 [hep-th]}{{\tt arXiv:0706.3359
  [hep-th]}}.

\bibitem{Freed:1991wd}
D.~S. Freed and R.~E. Gompf, ``{Computer calculation of Witten's three manifold
  invariant},'' {\em Commun. Math. Phys.} {\bf 141} (1991)
79--117.

\bibitem{CLM2011}
A.~Castro, N.~Lashkari, and A.~Maloney, {\em to appear}.

\bibitem{Maloney:2009ck}
A.~Maloney, W.~Song, and A.~Strominger, ``{Chiral Gravity, Log Gravity and
  Extremal CFT},'' {\em Phys. Rev.} {\bf D81} (2010) 064007,
\href{http://arXiv.org/abs/0903.4573}{{\tt 0903.4573}}.

\bibitem{Gibbons:1977mu}
G.~W. Gibbons and S.~W. Hawking, ``{Cosmological Event Horizons,
  Thermodynamics, and Particle Creation},'' {\em Phys. Rev.} {\bf D15} (1977)
2738--2751.

\bibitem{Park:1998qk}
M.-I. Park, ``{Statistical entropy of three-dimensional Kerr-de Sitter
  space},'' {\em Phys.Lett.} {\bf B440} (1998) 275--282,
  \href{http://arXiv.org/abs/hep-th/9806119}{{\tt hep-th/9806119}}.

\bibitem{Bousso:2001mw}
R.~Bousso, A.~Maloney, and A.~Strominger, ``{Conformal vacua and entropy in de
  Sitter space},'' {\em Phys. Rev.} {\bf D65} (2002) 104039,
\href{http://arXiv.org/abs/hep-th/0112218}{{\tt hep-th/0112218}}.

\bibitem{Balasubramanian:2002zh}
V.~Balasubramanian, J.~de~Boer, and D.~Minic, ``{Exploring de Sitter space and
  holography},'' {\em Class. Quant. Grav.} {\bf 19} (2002) 5655--5700,
\href{http://arXiv.org/abs/hep-th/0207245}{{\tt hep-th/0207245}}.

\bibitem{Quine}
J.~Quine and J.~Choi, ``{Zeta Regularized Products and Functional Determinants
  on Spheress},'' {\em Rocky Moun. J. Math.} {\bf 26} (1996) 719.

\bibitem{Iwaniec:2002aa}
H.~Iwaniec, {\em {Spectral methods of automorphic forms}}, vol.~53 of {\em
  Graduate Studies in Mathematics}.
\newblock American Mathematical Society, Providence, RI, second~ed., 2002.

\end{thebibliography}\endgroup

\end{document}